\documentclass[journal]{IEEEtran}
\usepackage{amsmath}
\usepackage{amssymb}
\usepackage{theorem}
\usepackage{mathrsfs} %pour mathscr
\usepackage[usenames,dvipsnames]{pstricks}
\usepackage{euscript}
\usepackage{graphicx}
\usepackage{enumitem}
\usepackage{charter}
\usepackage[english]{babel}
\usepackage{bm}
\usepackage{footnote}
\usepackage{lipsum}
\usepackage{multicol}
\newcommand{\email}[1]{\href{mailto:#1}{\nolinkurl{#1}}}
\definecolor{labelkey}{rgb}{0,0.08,0.45}
\definecolor{refkey}{rgb}{0,0.6,0.0}
\definecolor{Brown}{rgb}{0.45,0.0,0.05}
\definecolor{dgreen}{rgb}{0.00,0.49,0.00}
\definecolor{dblue}{rgb}{0,0.08,0.75}
\usepackage{cleveref}
\usepackage{multirow}

\newcommand{\bb}{\boldsymbol}
\renewcommand{\arraystretch}{1.0}
\PassOptionsToPackage{normalem}{ulem}
\usepackage{ulem}

\renewcommand{\le}{\ensuremath{\leqslant}}
\renewcommand{\ge}{\ensuremath{\geqslant}}
\newcommand{\minimize}[2]{\ensuremath{\underset{\substack{{#1}}}%
{\text{\rm minimize}}\;\;#2}}

\newcommand{\menge}[2]{\big\{{#1}~\big |~{#2}\big\}}

\newcommand{\HH}{\ensuremath{{\mathcal H}}}

\newcommand{\emp}{\ensuremath{{\varnothing}}}

\newcommand{\ID}{\boldsymbol{\mathsf{Id}\,}}

\newcommand{\RR}{\ensuremath{\mathbb{R}}}
\newcommand{\RP}{\ensuremath{\left[0,+\infty\right[}}

\newcommand{\RPP}{\ensuremath{\left]0,+\infty\right[}}

\newcommand{\RX}{\ensuremath{\left]-\infty,+\infty\right]}}

\newcommand{\pinf}{\ensuremath{{+\infty}}}

\newcommand{\dom}{\ensuremath{\text{\rm dom}\,}}
\newcommand{\prox}{\ensuremath{\text{\rm prox}}}

\newcommand{\argmind}[2]{\ensuremath{\underset{\substack{{#1}}}%
{\text{\rm argmin}}\;\;#2 }}

\newtheorem{theorem}{Theorem}[section]

\newtheorem{proposition}[theorem]{Proposition}
\theoremstyle{plain}{\theorembodyfont{\rmfamily}%
}
\theoremstyle{plain}{\theorembodyfont{\rmfamily}%
}
\theoremstyle{plain}{\theorembodyfont{\rmfamily}%
}
\theoremstyle{plain}{\theorembodyfont{\rmfamily}%
}
\theoremstyle{plain}{\theorembodyfont{\rmfamily}%
\theoremstyle{plain}{\theorembodyfont{\rmfamily}%
\newtheorem{remark}[theorem]{Remark}}
\theoremstyle{plain}{\theorembodyfont{\rmfamily}%
}
\theoremstyle{plain}{\theorembodyfont{\rmfamily}%
}
\renewcommand\theenumi{(\roman{enumi})}
\renewcommand\theenumii{\alph{enumii})}

\newcommand{\sm}{\mathcal{S}_K}
\newcommand{\smsp}{\mathcal{S}_K^{+}}
\newcommand{\smp}{\mathcal{S}_K^{++}}

\usepackage{amsmath}
\usepackage{amssymb}
\newcommand{\vectorize}[1]{\mathbf{#1}}
\newcommand{\vx}{\vectorize{x}}
\newcommand{\vy}{\vectorize{y}}

\newcommand{\vC}{\vectorize{C}}

\newcommand{\vQ}{\vectorize{Q}}

\newcommand{\vU}{\vectorize{U}}

\newcommand{\vd}{\vectorize{d}}
\newcommand{\vp}{\vectorize{p}}

\newcommand{\vu}{\vectorize{u}}
\newcommand{\vv}{\vectorize{v}}
\newcommand{\vm}{\vectorize{m}}
\newcommand{\vmu}{\boldsymbol{\mu}}

\newcommand{\vzero}{\vectorize{0}}

\newcommand{\vtau}{\boldsymbol{\tau}}
\newcommand{\vthe}{\boldsymbol{\theta}}

\DeclareMathOperator{\Diag}{Diag}

\usepackage{times}

\begin{document}

% Title.
% ------
% \title{Robust Estimation of Regularized Multivariate Exponential Power Models \NO{Convex?}}
\title{Convex Parameter Estimation of Perturbed Multivariate Generalized Gaussian Distributions}

% Authors
% -------
\author{Nora~Ouzir,~\IEEEmembership{Member,~IEEE,}
        Frédéric~Pascal,~\IEEEmembership{Senior member,~IEEE,}
        and~Jean-Christophe~Pesquet,~\IEEEmembership{IEEE Fellow member}% <-this % stops a space
\thanks{Nora Ouzir and Jean-Christophe Pesquet are with Université Paris-Saclay, CentraleSupélec, Inria OPIS, Centre de Vision Numérique, 91190, Gif-sur-Yvette, France. Frédéric Pascal is with Université Paris-Saclay, CNRS, CentraleSupélec, Laboratoire des Signaux et Systèmes, 91190, Gif-sur-Yvette, France.
The work of Jean-Christophe Pesquet was supported by BRIDGEABLE ANR Chair in AI. DGA has partially funded the research of Fr\'ed\'eric Pascal under grant ANR-17-ASTR-0015.}}
\markboth{Journal of \LaTeX\ Class Files,~Vol.~18, No.~9, September~2020}%
{How to Use the IEEEtran \LaTeX \ Templates}
\maketitle
% Abstract
% -------
\begin{abstract}
The multivariate generalized Gaussian distribution (MGGD), also known as the multivariate exponential power (MEP) distribution, is widely used in signal and image processing. %\NO{It can model a broad range of signals.} 
However, estimating MGGD parameters, which is required in practical applications, still faces specific theoretical challenges. In particular, establishing convergence properties for the standard fixed-point approach when both the distribution mean and the scatter (or the precision) matrix are unknown is still an open problem. In robust estimation, imposing classical constraints on the precision matrix, such as sparsity, has been limited by the non-convexity of the resulting cost function.
This paper tackles these issues from an optimization viewpoint by proposing a convex formulation with well-established convergence properties. We embed our analysis in a noisy scenario where robustness is induced by modelling multiplicative perturbations.
%A re-parametrization of the original likelihood function is at the heart of the proposed approach. 
The resulting framework is flexible as it combines a variety of regularizations for the precision matrix, the mean and model perturbations. This paper presents proof of the desired theoretical properties, specifies the conditions preserving these properties for different regularization choices and designs a general proximal primal-dual optimization strategy. 
The experiments show a more accurate precision and covariance matrix estimation with similar performance for the mean vector parameter compared to Tyler's $M$-estimator. In a high-dimensional setting, the proposed method outperforms the classical GLASSO, one of its robust extensions, and the regularized Tyler's estimator.
% Compared with Tyler's M-estimator, the proposed method shows a more accurate precision and covariance matrix estimation with similar performance for the mean vector parameter. In a high-dimensional setting, the proposed method outperforms the classical GLASSO and its robust extension, as well as the regularized Tyler's estimator.}
%
%% More precisely, in noisy scenarios, the robust estimation of the MGGD parameters has been traditionally addressed by a fixed-point approach associated with a nonconvex optimization problem.  
% As an alternative, this paper presents a novel convex formulation for robustly estimating MGGD parameters in the presence of multiplicative perturbations. The proposed approach is grounded on a re-parametrization of the original likelihood function in a way that ensures convexity. We also show that this property is preserved for several typical regularization functions. 
%\NO{For real datasets, the proposed method shows promising results in clustering tasks...}
\end{abstract}
\IEEEpeerreviewmaketitle

% Intro
% ------
\section{Introduction}
\label{sec:intro}
\IEEEPARstart{P}{probabilistic} distribution parameters, particularly the mean and covariance  --or scatter/precision matrix--, are central in statistical signal processing. In machine learning applications, they are building blocks of clustering, classification, dimension reduction, or detection models \cite{roizman2019flexible, houdouin2022robust, zhang2014novel}. These parameters are rarely known in practice, and estimating them has been a long-standing statistical problem. A Maximum Likelihood (ML) approach relying on the standard Gaussian assumption has long been the solution of choice, bringing simplicity and tractability to performance analysis. The statistical properties of the Gaussian ML estimators are also well-known: the estimated mean is Gaussian, while the covariance matrix estimator follows a Wishart distribution. 
%
% In our time, however, real datasets are increasingly challenging this simplistic Gaussian assumption. Not only are they more and more massive and high-dimensional, but they are frequently heterogeneous and perturbed by significant outliers. In various domains of signal processing, such as radar~\cite{de2015modern}, financial signal~\cite{feng2016signal}, and image processing~\cite{Pasc13}, these limitations are well-known.
% The widely used Gaussian probabilistic model tends to perform poorly in these areas, which has led to the search for more suitable alternatives.

The standard Gaussian assumption is no longer sufficient to handle the complexity of real-world datasets today. With datasets becoming increasingly large, high-dimensional, and heterogeneous, and often containing significant outliers, the limitations of the Gaussian probabilistic model are becoming more apparent. In domains such as radar~\cite{de2015modern}, financial signal~\cite{feng2016signal}, and image processing~\cite{Pasc13}, where these limitations are well-known, alternative models are being sought to address these challenges.
Robust estimation theory is one of those alternatives that can provably deal with data perturbations~\cite{maronna1976robust}. It models observations with elliptically-contoured (EC) (complex and real Elliptically Symmetric (ES)) distributions. This broad-ranging model encompasses well-known multivariate distributions such as the Gaussian, Generalized Gaussian, $t$-, or $k$-distributions (see~\cite{ollila2012complex} for a review). %\NO{\textbf{Link to MGGD here, explain the need to establish convergence properties in the case where the mean is unknown. state contribution clearly?.}}

Regularizing the parameter estimation problem has been another way of tackling current data challenges, such as high dimensionality. For example, sparse $\ell_1$-regularization of the precision matrix is a popular choice for Gaussian distributions~\cite{GLASSO08}. Works in the standard Gaussian framework have also laid the grounds for regularizing robust estimators of EC models. In~\cite{ollila2014regularized}, the scatter matrix has been regularized through the trace of its inverse. At the same time, closely related shrinkage Tyler's estimators have been studied in~\cite{pascal2014generalized,sun2014regularized}. Although these works have addressed the problem in the general statistical context of EC models, they remain limited from the regularization viewpoint.
%Although addressing the problem in the general statistical context of EC models, these works have considered only particular regularization approaches. 
For example, unlike existing works in the Gaussian case, they have not considered $\ell_1$-regularization of the precision matrix due to the non-convexity of the resulting problem in robust estimation. It is also worth noting that mostly plug-in (as opposed to model-based) robust estimation strategies have been proposed in the Gaussian case with $\ell_1$-regularization~\cite{Lafit2023, tarr2016robust, Ollerer2015}. In \cite{wiesel2012geodesic}, a framework based on geodesic convexity has been proposed to regularize the perturbed covariance matrix of a zero-mean Gaussian distribution. Furthermore, previous works have not explored regularizing other parameters, such as the mean or data perturbations. As explained in the following, this work proposes an original model-based approach that copes with the non-convexity of the original problem and provides a general framework for exploring a larger variety of regularizations, along with associated theoretical conditions.%

This work tackles the above issues for a popular subclass of EC models: the MGGD~\cite{Pasc13}, also called MEP distribution~\cite{gomez1998multivariate}. This distribution family is instrumental for image processing because of its ability to model a wide range of features (\textit{e.g.}, wavelet coefficients, gradients)~\cite{Pasc13, kumar2020semi}. It is also increasingly used to model weight distributions in (deep) neural networks~\cite{wang2019deep, chai2019using}. From a statistical point of view, its much-needed flexibility is obtained thanks to an extra shape parameter $\beta$ modelling lighter ($\beta>2$) and heavier than the Gaussian ($\beta<2$) tails.\footnote{Note that $\beta$ in this work corresponds to $2\beta$ in most MGGD-related literature. This choice allows us to simplify notations.} The multivariate Gaussian distribution is a special case of MGGDs when $\beta = 2$. Furthermore, the parameter estimation problem is addressed in this work in the particular context of perturbed MGGDs. We are specifically interested in jointly estimating the mean vector, covariance matrix, and perturbation parameters, a complex problem introduced in \cite{ouzir2022convex}\footnote{A preliminary version of this work was presented in \cite{ouzir2022convex} with
reduced theoretical results and limited experimental validation.}, with broad applicability in real-world applications. Our main contribution is a convex reformulation of the resulting statistically robust joint parameter estimation problem. The proposed formulation offers all estimators theoretical guarantees of existence and establishes conditions for their uniqueness. In addition to modelling heterogeneity and outliers, the proposed formulation is flexible and can adapt to various real-life data structures. For example, the sparsity of the precision matrix can be incorporated through $\ell_1$- regularization to deal with high dimensional data with relatively few observations. Unlike previous approaches, the proposed framework enables regularizing all the involved parameters beyond the precision matrix. We also study the impact of these regularizations and specify the theoretical conditions they must satisfy to preserve convexity. Our second contribution is designing a proximal primal-dual algorithm tailored to solve the parameter estimation problem with guaranteed convergence~\cite{Komodakis2015, combettes2021fixed}. The proposed algorithm is flexible, and associated proximity operators can easily incorporate various regularization choices. 

The paper is organized as follows. Section~\ref{sec:2} introduces the proposed perturbed MGGD model, starting with the underlying statistical framework. Section~\ref{sec:3} presents the proposed convex parameter estimation approach, which is the main contribution of this paper. The proposed proximal primal-dual algorithm is detailed in Section~\ref{se:PFalgo}. In Section~\ref{sec:4}, we compare the performance of the proposed approach to different state-of-the-art robust and non-robust estimators in various experimental scenarios. Finally, concluding remarks and perspectives are provided in Section~\ref{sec:conclusion}.\\

\noindent \textbf{Notations}: $\sm$ denotes the space of symmetric real matrices of size $K\times K$, $\smsp$ is the cone of positive semi-definite matrices, and $\smp$ the cone of positive definite matrices. $\ID$ denotes the identity matrix (whatever its size), $\boldsymbol{1}$ the vector (the dimension of which is understood from the context) whose components are all equal to 1, and $\rm tr(.)$ the trace of a matrix. $\|\cdot\|$ denotes the Euclidean norm, $\|.\|_{\rm S}$ is the operator norm, and $\|\cdot\|_{r}$ with $r\in [1,+\infty[$ denotes the element-wise $\ell^r$ norm (the same notation will be used for a matrix or for a vector whatever the dimension). $\Gamma_{0}(\HH)$ denotes the class of lower-semicontinuous convex functions from some Hilbert space $\HH$ to $\RX$ which are proper (\textit{i.e.}, finite at least at one point). The domain of $f$ is $\dom f = \menge{x\in \HH}{f(x)<\pinf}$. $\iota_{\mathcal{D}}$ denotes the indicator function of $\mathcal{D}\subset \HH$, which is equal to $0$ on this set and $\pinf$ out of it. Finally, a function $f\colon \HH \to \RX$ is coercive if $\lim_{\|x\|\to \pinf} f(x) = \pinf$.
%
% Problem
% -------
%
%\newpage 
\section{Problem Formulation}\label{sec:2}
The MGGD~\cite{gomez1998multivariate} belongs to a broad subclass of EC distributions that can model various uni-modal probability density functions (p.d.f) with heavier or lighter tails than the Gaussian one. A random vector following an EC distribution can be defined using its stochastic representation \cite{yao1973representation}:
\begin{equation}
\label{stochastic-representation}
\mathbf{x} \overset{d}{=} \boldsymbol{\mu} + \mathcal R \, \mathbf{A} \, \mathbf{u},
\end{equation}
where $\overset{d}{=}$ stands for ``is distributed as''. The \textit{modular variate} $\mathcal R$ is a positive random variable (with an unknown p.d.f), $\mathbf{u}$ is a random vector uniformly distributed on the unit-hypersphere $\{\mathbf{u} \in \mathbb R^K | \, \| \mathbf{u} \|=1\}$, with $\mathcal R$ and $\mathbf{u}$ (statistically) independent. In \eqref{stochastic-representation}, the mean vector $\boldsymbol{\mu}$ and the scatter matrix factorization $\mathbf A$ (and the associated matrix $\mathbf C = \mathbf A\, \mathbf A^\top$) 
%\textcolor{orange}{JCP: avec les hypothèses faites sur  $\mathbf{u}$, cela ne me paraît pas clair que c'est la même matrice $\mathbf C$ que dans la suite.} \FP{Et pourtant c'est le cas.}
are the unknown parameters of the EC model. The stochastic representation described above is of practical interest for simulating EC-distributed random vectors. In particular, for the MGGD subclass,  $\mathcal R^\beta \sim \Gamma(1/2, K/\beta)$, where $\Gamma(a,b)$ is  the  univariate  Gamma  distribution  with  parameters $a$ and $b$ (see \cite{johnson1995continuous} for a definition).\\

The zero-mean MGGD of dimension $K$, denoted by MGGD$_K(\beta, \mathbf 0,\vC)$, is characterized by its p.d.f
\begin{equation}\label{eq:pdf}
%\resizebox{0.5\textwidth}{!}{%
\mathsf{p}_{\mathsf{x}}(\cdot) = C_{K,\beta}(\det \vC)^{-1/2} \exp\!\left(\!-\frac12 \Big[(\cdot)^\top \!\vC^{-1}\! (\cdot)\Big]^{\beta/2}\!\right)\!,%
\end{equation}
where $C_{K,\beta} = \cfrac{K \, \Gamma(K/2)}{\pi^{K/2}\Gamma(1\!+\!K/\beta)2^{1+K/\beta}}$ is a constant, $\Gamma(\cdot)$ denotes the Gamma function, and $\vC\in \smp$ is (up to a multiplicative factor) the associated covariance matrix.\footnote{The covariance matrix of the MGGD is equal to $2^{2/ \beta} \Gamma ((K+2)/ \beta )(K \, \Gamma (K/ \beta ))^{-1} \,\mathbf{C}$. For large values of $K$, the following approximation can be used (for computational purposes): $\Gamma(a\,K+b) \sim \sqrt{2\,\pi} \exp(-a\,K) (a\,K)^{a\,K+b-1/2}$, leading to $2^{2/ \beta} \Gamma ((K+2)/ \beta )(K \, \Gamma (K/ \beta ))^{-1} \,\mathbf{C} \sim 1/K \,(2\,K/\beta)^{2/\beta}\,\mathbf{C}$.} 
%Note again that, in this work, we use $\beta/2$ instead of $\beta$ to simplify notations.  
With this notation, when $\beta = 2$, \eqref{eq:pdf} corresponds to the standard multivariate Gaussian distribution. When $\beta < 2$, distributions with heavier tails than the Gaussian one are obtained. In the remainder of this paper, we assume that the exponent $\beta> 1$ is known (see Section~\ref{sec:shape_parameter}).
%
% ----------------------------
\subsection{Observation Model with Multiplicative Perturbations}
% ----------------------------
%
This subsection introduces the proposed MGGD model under multiplicative perturbations. Let $(\vx_{n})_{1\le n\le N}$ be $N$ realizations of independent and identically distributed (\textit{i.i.d.}) random $K$-dimensional vectors generated according to a zero-mean MGGD with shape parameter $\beta$. Let us now consider a scenario where noisy observations $(\vy_{n})_{1\le n\le N}$ result from a multiplicative perturbation $\vtau$ of the unperturbed $(\vx_{n})_{1\le n\le N}$. The corresponding observation model reads
\begin{equation}\label{e:statmodel}
(\forall n \in \{1,\ldots,N\})
\quad \mathbf{y}_{n} = \tau_{n}
%\theta_{n}^{1-1/\beta} 
\mathbf{x}_{n}+\boldsymbol{\mu},
\end{equation}
where $\vtau = (\tau_{n})_{1\le n \le N}\in \RPP^N$ and $\beta\in ]1,+\infty[$. The $\vtau$-perturbed model \eqref{e:statmodel} is equivalent to the assumption that the $\vy_n$ samples follow a MGGD$_K(\beta, \boldsymbol{\mu}, \tau_n^2\,\vC)$.  
As \eqref{e:statmodel} accounts for possible outliers in the observations (\textit{i.e.,} $\tau_n$ values larger than $1$), it can also be seen as a general EC distribution where $\tau_n$'s are realizations of an unknown positive p.d.f. Interestingly, similar models have been widely studied in the particular context of perturbed Gaussian distributions (see, \textit{e.g.}, \cite{pascal2007covariance}).  

\noindent Starting from \eqref{e:statmodel}, we aim to jointly estimate the noise and the unknown distribution parameters $\vmu$ and $\vC$. Thus, $N$ scalar parameters $(\tau_n)_{1\le n\le N}$ need to be estimated in addition to the mean and covariance matrix. This problem has been addressed in~\cite{pascal2007covariance} for the particular case of a centred Gaussian distribution ($\vmu = \mathbf 0$ and $\beta=2$.) The following subsection recalls the existing ML-based approaches in the more general case.
%
%------------------------------------
\subsection{Previous Work on Estimating $(\vC,\vmu,\vtau)$}
%------------------------------------
%
ML estimation of the unknown parameters $(\vC,\vmu,\vtau)$ has been broadly studied for EC models, including the MGGD or compound Gaussian models. (For non-perturbed EC models, one can refer to \cite{maronna1976robust, chitour2008exact}.) For the proposed $\vtau$-perturbed MGGD model, estimating $\vC$, $\vmu$, and $\vtau$ can be achieved by minimizing the negative log-likelihood function arising from~\eqref{e:statmodel} (up to the normalizing constant that does not depend on the unknown parameters), \textit{i.e.},
% The resulting negative log-likelihood for model \eqref{e:statmodel} thus reads
%
\begin{multline}
\label{e:nLL}
\mathcal{L}(\vC,\vmu,\vtau) = 
\frac{1}{2}\! \sum_{n=1}^N \frac{\big[(\vy_{n}\! -\! \vmu)^\top \vC^{-1} (\vy_{n}\! -\! \vmu)\big]^{\beta/2}}{\tau_{n}^{\beta}} \\
+ \frac{N}{2}\! \log \det \vC + K\!  \sum_{n=1}^{N} \log \tau_{n}.
\end{multline}
Function~\eqref{e:nLL} is non-convex, and its minimization has been intensively investigated~\cite{Chen11, Pasc13}: the standard approach being to first minimize with respect to $\vtau$, then plug the optimal value $\widehat{\vtau}(\vC,\vmu)$ into~\eqref{e:nLL}. Precisely, the first step of the standard approach yields % for $\vtau$: %so yielding the optimal value
\begin{equation}\label{eq:tauhat}
\widehat{\vtau}(\vC,\vmu) =\left[\Big(\frac{ \beta [(\vy_{n}-\vmu)^\top \vC^{-1} (\vy_{n}-\vmu)]^{\beta/2} }{2 K}\Big)^{\frac{1}{\beta}}\right]_{1\le n \le N},
%\frac{1}{K} (\vy_{n}^\top \vC^{-1} \vy_{n})_{1\le n \le N}.
\end{equation}
and by assuming none of the vectors $(\vy_{n})_{1\le n \le N}$ is equal to $\vmu$ (which is true with probability $1$ for continuous random vectors for a given $\vmu$), plugging \eqref{eq:tauhat} into \eqref{e:nLL} leads to
\begin{multline}\label{eq:fpoint}
\mathcal{L}(\vC,\vmu,\widehat{\vtau}(\vC,\vmu))  = \frac{K}{2} \!\sum_{n=1}^{N} \log\big[(\vy_{n}\!-\!\vmu)^\top \vC^{-1} (\vy_{n}\!-\!\vmu)\big]\\
+\frac{N}{2}\!\log \det \vC, %+.
\end{multline}
where the constant term $\frac{KN}{\beta} \!\left[1\!+\! \log\Big(\frac{2K}{\beta}\Big)\right]$ has been omitted. Note that \eqref{eq:fpoint} is (up to constant terms) the log-likelihood function of the (central) angular Gaussian distribution~\cite{tyler1987statistical}. For a given value of $\vmu$, minimizing \eqref{eq:fpoint} with respect to $\vC$ (or $\vC^{-1}$) on $\smp$ thus leads to Tyler's estimator \cite{tyler1987distribution}, defined as the unique (up to a scale factor) solution to the fixed-point equation:
\begin{equation}
    \label{Tylers-est}
    \vC = \cfrac{K}{N} \sum_{n=1}^N \cfrac{(\vy_{n}\!-\!\vmu)(\vy_{n}\!-\!\vmu)^\top}{(\vy_{n}\!-\!\vmu)^\top \vC^{-1} (\vy_{n}\!-\!\vmu)}.
\end{equation}
Considering both $\vC$ and $\vmu$ unknown leads to the joint fixed-point equations~\cite{frontera2015hyperspectral}:
\begin{equation}
    \label{Tylers-est-mu-C}
    \left\{
    \begin{array}{cc}
       \vmu  = & \cfrac {\displaystyle \sum_{n=1}^{N} \,\cfrac{\vy_{n}}{(\vy_{n}\!-\!\vmu)^\top \vC^{-1} (\vy_{n}\!-\!\vmu)^{}}}{\displaystyle \sum_{i=1}^{N} \,\cfrac{1}{(\vy_{n}\!-\!\vmu)^\top \vC^{-1} (\vy_{n}\!-\!\vmu)^{}}} \, ,\\
       %\label{Tylers-est}
       \vC   = & \cfrac{K}{N} \displaystyle \sum_{n=1}^N \cfrac{(\vy_{n}\!-\!\vmu)(\vy_{n}\!-\!\vmu)^\top}{(\vy_{n}\!-\!\vmu)^\top \vC^{-1} (\vy_{n}\!-\!\vmu)}\,. 
    \end{array}
    \right.
\end{equation}
%\FP{Réponse page 245 du papier fondateur , qui définit le Tyler's M-est comme un Huber-type est. En revanche, si on dérive directement la vraisemblance Eq.(6), je suis d'accord qu'il n'y a pas de 1/2... Vous voulez mettre quoi? Perso, je pense que c'est mieux de mettre avec le 1/2 car c'est à ça qu'on se compare, mais il faut réécrire la phrase. ton avis JCP ?}
%
Tyler's estimator has been extensively studied from a computational and statistical perspective (\cite{tyler1987distribution, pascal2007covariance, pascal2008performance}). Note that when one redefines Tyler's estimator as a Huber-type estimator~\cite{tyler1987distribution}, an exponent $1/2$ is added to the denominator in \eqref{Tylers-est-mu-C}. The following remarks are of importance. 
\begin{remark}
On the joint fixed-point algorithm:
\begin{enumerate}
    \item First, the existence, uniqueness, and convergence of the recursive algorithm associated with \eqref{Tylers-est-mu-C} \textbf{has yet to be proved}, although these estimators have been introduced in the 80s~\cite{tyler1987distribution}. Establishing convergence properties when both $\vmu$ and $\vC$ are unknown is still an open issue.
    If existence/uniqueness is assumed, then the asymptotic joint distribution of the corresponding estimators $(\hat{\vmu},\hat{\vC})$ is entirely characterized \cite{tyler1987distribution}.
%\end{itemize}
%\end{remark}
%
%\begin{remark}
\item The resulting estimators $(\hat{\vmu},\hat{\vC})$ \textbf{do not depend on} $\beta$. It can be shown (see \cite{roizman2019flexible}) that this is even true for any $\vtau$-perturbed EC distribution expressed by \eqref{e:statmodel}, where \textit{extra}-parameters, such as shape, degree of freedom, or scale, only impact the estimation of $\vtau$. 
\end{enumerate}
\end{remark}
In this work, we prove the existence, uniqueness, and convergence of all three $\vmu$, $\vC$, and $\vtau$ estimators in a regularized setting. %Furthermore, we recover that $\hat{\vmu}$ and $\hat{\vC}$ do not depend on $\beta$.
%\textcolor{orange}{Je ne trouve pas cela évident à partir de notre approche. Si ce n'est pas évident, mieux vaut éviter la remarque précédente et garder cela pour un travail ultérieur. Ou peut-être peut-on reprendre une partie de cette remarque au début du paragraphe suivant?} \FP{Il me semblait que c'était clair dans l'algo d'optim car less équations pour obtenir $\vC$ et $\mu$ ne dépendent pas de $\beta$, non ? On peut toujours le mettre dans la section suivante.} \NO{Je dis peut-etre nimporte quoi mais: dans les updates de l'algo il y a un lien a travers le $\prox_{\gamma \varphi}$qui va lier u et $\theta_1$ qui lui meme est en lien avec $\beta$ (voir la solution de lequation (77) ($t(\vu,\xi)$))}.% Si on suit la meme logique de dire que le $\theta_1$ (tau) depend de $\beta$ a literation davant et est fixé ensuite alors il n'y a plus de dependence si: la solution de lequation (77) ($t(\vu,\xi)$) ne depend pas de $\beta^*$.}

%
\subsection{Shape Parameter $\beta$}~\label{sec:shape_parameter}
As mentioned previously, the shape parameter $\beta$ only plays a role in estimating the perturbation $\vtau$. One could compensate for an approximate $\beta$ value and retrieve the original distribution by varying $\vtau$, regardless of the true shape of the MGGD. Furthermore, estimating the (\textit{extra}-)shape parameter of the EC model is a difficult problem that has been studied intensively in the non-perturbed MGGD case~\cite{sharifi1995estimation, song2006globally, dominguez2003practical, krupinski2006approximated}. Other \textit{extra}-parameters, such as the degree of freedom of the $t$-distribution~\cite{pascal2021improved} or the shape parameter of a $k$-distribution~\cite{abraham2010reliable} have also been studied. 

With this in mind, estimating $\beta$ for the presented $\vtau$-perturbed MGGD model appears out of the scope of this work. We will assume that the shape parameter is known or previously estimated in a non-perturbed MGGD context for the experiments.
%
%------------------------------------
\section{Proposed Convex Formulation}\label{sec:3}
%------------------------------------
%
This section introduces a convex alternative to minimizing \eqref{e:nLL}. The main idea of the proposed approach is to convexify the cost function through suitable variable changes and regularization. The purpose of regularizing the cost function is two-fold; first, prior information about the sought variables may be used to improve their estimation (\textit{e.g.}, sparsity of the precision matrix $\vC^{-1}$~\cite{GLASSO08}). Secondly, regularization is critical in tackling the non-convexity of the cost function. The following subsection explains the details of these transformations. 
\subsection{A Regularized Cost Function}
%--------------------------------------
As explained above, convexifying \eqref{e:nLL} requires a series of transformations. First, let us re-parameterize $\mathcal{L}$ by setting 
\begin{align}
\vQ & =\vC^{-1/2}\label{eq:reparam_Q}\\%\vC^{-1} &= \vQ^2\\
\vm & = \vQ \vmu\label{eq:reparam_m}\\
\vthe & = (\theta_{n})_{1\le n \le N}= (\tau_{n}^{\beta/(\beta-1)})_{1\le n \le N}\label{eq:reparam_theta},
\end{align}
where $\vQ\in \smp$. Using the re-parametrization \eqref{eq:reparam_Q}-\eqref{eq:reparam_theta}, we can rewrite the original cost function as $\mathcal{L}(\vC,\vmu,\vtau) = \widetilde{\mathcal{L}}(\vQ,\vm,\vthe)$, where
\begin{multline}\label{eq:newL}
% \resizebox{0.5\textwidth}{!}{%
\widetilde{\mathcal{L}}(\vQ,\vm,\vthe)
=\frac12 \!\sum_{n=1}^N \frac{\| \vQ \vy_{n}-\vm\|^\beta}{\theta_{n}^{\beta-1}}
- N\!\log \det \vQ  \\+ K(1\!-\!1/\beta)\! \sum_{n=1}^{N} \log \theta_{n}.
%}
\end{multline}
If we exclude the last term, the obtained function becomes convex with respect to $(\vQ,\vm,\vthe)$. In Subsections~\ref{sec:study_cost_fct} and \ref{sec:choice_of_reg_theta}, 
%we show that carefully choosing a regularization function for $\vthe$ will allow us to deal with the non-convexity of the last term and recover a fully convex formulation.
we show that carefully choosing a regularization function for $\vthe$ can counteract the concavity of the last term in \eqref{eq:newL} and lead to a fully convex formulation. In the following, we first introduce the general regularized form of \eqref{eq:newL} that accounts for prior information on the variables:
%
% %--------------------------------------
% \subsection{Regularization}
% % -------------------------------------
% % 
%To account for this prior information, we also introduce a regularized cost function 
%--
\begin{multline}\label{e:costfunc}
f(\vQ,\vm,\vthe) = \\
\resizebox{0.49\textwidth}{!}{%
$
\begin{cases}
\widetilde{\mathcal{L}}(\vQ,\vm,\vthe,\vd)\!+\!g_{\mathsf{Q}}(\vQ)\!+\!g_{\mathsf{m}}(\vm)\!+\!g_{\vartheta}(\vthe)
&\!\mbox{if $\vthe \in \RPP^N$}\\
&\!\mbox{and $\vQ \in \smp$}\\
\pinf &\!\mbox{otherwise,}
\end{cases}
$%
}
\end{multline}
%\FP{On ne met pas $g_{{/vm}}$ au lieu de $g_{{m}}$? Idem pour les autres...}\NO{NO: a mon avis c'est plus lisible/leger sans le gras}\\
%\FP{Il y a un $\vartheta$ qui apparait ici sans explication (même remarque que précédemment).}\\
%--
where $g_{\mathsf{Q}}\colon \sm\to \RX$, $g_{\mathsf{m}}\colon \RR^K \to \RX$, and $g_{\vartheta}\colon \RR^N\to \RX$ are regularization functions on $\vQ$, $\vm$, and $\vthe$, respectively.\footnote{These functions can take infinity values to model potential hard constraints on the variables. For example, if one seeks to restrict the vector $\vm$ to some set $\mathcal{D} \subset \RR^K$ (\textit{e.g.}, some hypercube or some ball) a suitable choice for $g_{\mathsf{m}}$ is the indicator function $\iota_{\mathcal{D}}$ of $\mathcal{D}$, equal to 0 on $\mathcal{D}$ and $\pinf$ elsewhere.} From a Bayesian perspective, minimizing the regularized cost function \eqref{e:costfunc} amounts to a Maximum A Posteriori (MAP) estimation of $\vQ$, $\vm$, and $\vthe$. The regularization functions introduced in \eqref{e:costfunc} can be viewed as the potentials associated with (possibly improper) prior probability density functions proportional to $\exp(-g_{\mathsf{Q}}(\cdot))$, $\exp(-g_{\mathsf{m}}(\cdot))$, and $\exp(-g_{\vartheta}(\cdot))$, respectively. %Minimizing \eqref{e:costfunc} then amounts to computing Maximum A Posteriori (MAP) estimates of $\vQ$, $\vm$, and $\vthe$. The choice of these regularization functions will be discussed in detail in Section~\label{}.

The regularized cost function $f$ is suitable for different robust
estimation problems depending on the choice of the regularization functions. For example, in graph processing applications, it is known that the precision matrix $\vC^{-1}$ is sparse~\cite{GLASSO08}. Prior information on the nature of the perturbation $\vtau$ may also be accessible (\textit{e.g.,} bounds), or one may seek to restrict the mean to a specific set (\textit{e.g.,} known values for a restricted subset of components). Examples of different practical regularizations will be discussed in more detail in Subsections~\ref{sec:choice_of_reg_Qm} and \ref{sec:choice_of_reg_theta}. 
\subsection{Study of Cost Function $f$}\label{sec:study_cost_fct}
%------------------------------------------
%
This subsection discusses key properties of the proposed cost function $f$. We will rely on these properties to choose the regularization functions and develop a suitable minimization strategy. Namely, we introduce two central propositions describing the assumptions that will ensure convexity and the existence of a minimizer. (Proof outlines are provided in Subsection~\ref{sec:sketch_proof}, and detailed proofs of these propositions can be found in Appendices A--B.) To simplify the analysis, we start by introducing the following functions: 
\begin{align}
&\psi\colon \RR \to \RX\colon \xi \mapsto 
\begin{cases}
-\log \xi & \mbox{if $\xi > 0$}\\
\pinf & \mbox{otherwise,}
\end{cases}\label{eq:defpsi}\\
&\widetilde{g}_{\vartheta}\colon \RR^N \to \RX\colon \vthe \mapsto
%\begin{cases}
g_{\vartheta}(\vthe)  - K(1-1/\beta) \sum_{n=1}^{N} \psi(\theta_{n}),
%\log \theta_{n} & \mbox{if $\vthe \in \RPP^{N}$}\\
%\pinf & \mbox{otherwise,}
%\end{cases}
\label{e:deftildeg}\\
&\Psi\colon \sm \to \RX \colon \vQ \mapsto 
\begin{cases}
- N\log \det \vQ  & \mbox{if $\vQ\in \smp$}\\
\pinf & \mbox{otherwise.}\label{e:defPsi}
\end{cases}
\end{align}
We then have the following results.
%------------------------------------------------
\begin{proposition}\label{prop:convex}
Assume that $g_{\mathsf{Q}}\in \Gamma_{0}(\sm)$ and that there exists
an invertible matrix $\overline{\vQ}\in \smp$ such that $g_{\mathsf{Q}}(\overline{\vQ}) <+\infty$.
Assume that $g_{\mathsf{m}} \in \Gamma_{0}(\RR^K)$ and $\widetilde{g}_{\vartheta}\in \Gamma_{0}(\RR^N)$.
Then, $f$ is a proper lower-semicontinuous convex function on $\sm\times \RR^K\times \RR^N$.
\end{proposition}
\vspace{\baselineskip}
\begin{proof}
See Appendix~\ref{Apx:prop_convex}.    
\end{proof}
\begin{proposition}\label{prop:minim}
In addition to the conditions stated in Proposition~\ref{prop:convex}, let us make the following assumptions:
\begin{enumerate}
\item \label{a:pmini} $g_{\mathsf{m}}\ge 0$;
\item \label{a:pminii} $\widetilde{g}_{\vartheta}= \widetilde{g}_{\vartheta,0}+\widetilde{g}_{\vartheta,1}$ 
where $\widetilde{g}_{\vartheta,0} \in \Gamma_{0}(\RR^N)$, $\widetilde{g}_{\vartheta,1} \in \Gamma_{0}(\RR^N)$,
$\dom \widetilde{g}_{\vartheta,1} = \RPP^{N}$,  and
\begin{equation}
(\forall \check{\vthe} \in \RP^N\setminus \RPP^N)\quad \lim_{\vthe\to \check{\vthe}}  \widetilde{g}_{\vartheta,1}(\vthe) = \pinf;
\end{equation}
\item \label{a:pminiii} $\Psi+g_{\mathsf{Q}}$ and $\widetilde{g}_{\vartheta}$ are coercive functions.
\end{enumerate}
Then $f$ admits a minimizer. Such a minimizer is unique if $g_{\mathsf{Q}}$ and $\widetilde{g}_{\vartheta}$ are strictly convex.
\end{proposition}
\vspace{\baselineskip}
\begin{proof}
See Appendix~\ref{Apx:prop_minim}.    
\end{proof}\\

It stems from Propositions \ref{prop:convex} and \ref{prop:minim} that the convexity of $f$ and the existence of a minimizer depend on the choice of the regularizations. First, one can easily notice that the conditions on $g_{\mathsf{m}}$ and $g_{\mathsf{Q}}$ are satisfied by a wide range of typical regularization functions. Secondly, the assumptions on $\widetilde{g}_{\vartheta}$ are less usual but critical regarding the convexity of $f$. More precisely:
\begin{itemize}
\item The assumptions on $g_{\mathsf{m}}$ are mild. They are satisfied both when no information is available on parameter $\vm$ ($g_{\mathsf{m}} = 0$) or when this vector is known ($g_{\mathsf{m}} = \iota_{\{\overline{\vm}\}}$ with $\overline{\vm}\in \RR^K$). In practice, restricting the mean to a specific set (\textit{e.g.}, known values for a restricted subset of components) satisfies these conditions.
\item 
%Citations are missing hereafter: 
The assumptions on $g_{\mathsf{Q}}$ are rather classical. In particular, the core assumptions are satisfied by standard convex penalization promoting sparsity or group sparsity~\cite{GLASSO08}. The strict convexity assumptions can be satisfied by adding a quadratic term leading to an elastic net-like penalization~\cite{hastie2005}. Subsection~\ref{sec:choice_of_reg_Qm} will explicit some useful regularization functions for $\vQ$. 
\item The key conditions for convexity are related to the choice of $\widetilde{g}_{\vartheta}$. Note that this function contains the regularization that acts implicitly on the perturbation parameter $\vtau$. In Subsection~\ref{sec:choice_of_reg_theta}, we propose a regularization satisfying the required conditions for $\widetilde{g}_{\vartheta}$ and analyse its impact on estimating MGGD parameters.
\end{itemize}
%
%--------------------------------------
\subsection{Proof Outlines}\label{sec:sketch_proof}
%--------------------------------------
%
\subsubsection{Proof of Proposition~\ref{prop:convex}}
The proof first shows that $f$ is a sum of proper lower-semicontinuous convex functions, which makes it a lower-semicontinuous convex function. %For this purpose, we start by defining the following functions
% Let us define the following functions:
% \begin{align}\label{e:defPhi}
% &\Phi\colon (\RR^K)^N\times \RR^{\NO{K (N?)}}\colon ((\vv_{n})_{1\le n\le N},\vthe) \mapsto \sum_{n=1}^N \varphi(\vv_{n},\theta_{n}),\\
% &\varphi\colon \RR^K\times \RR \colon (\vu,\xi)\mapsto
% \begin{cases}
% \displaystyle \frac{\|\vu\|^\beta}{2\xi^{\beta-1}} & \mbox{if $\vu \neq \vzero$ and $\xi > 0$}\\
% 0 & \mbox{if $\vu = \vzero$ and $\xi=0$}\\
% \pinf & \mbox{otherwise,}
% \end{cases}
% \end{align}
% and the linear operator
% \begin{multline}
% \mathcal{T}\colon \sm \times \RR^{K} \to (\RR^K)^N\colon
% (\vQ,\vm) \mapsto  (\vQ \vy_{n}-\vm)_{1\le n \le N}.
% \end{multline}
% The considered cost function can then be reexpressed as
% \begin{multline}
% \label{e:costreexpressed}
% f(\vQ,\vm,\vthe) = \Phi\big(\mathcal{T}(\vQ,\vm),\vthe\big)+ \Psi(\vQ) 
% +g_{\mathsf{Q}}(\vQ)\\+g_{\mathsf{m}}(\vm)+\widetilde{g}_{\vartheta}(\vthe)
% \end{multline}
%
More precisely, because of the form of the first term in $\widetilde{\mathcal{L}}$, $f$ involves the lower-semicontinuous envelope of 
a perspective function~\cite{Com18}, which leads to the first part of the function $f$ being lower-semicontinuous and convex according to \cite[Proposition~9.42]{Livre1}. In addition, the assumptions made on the different regularizations in Proposition~\ref{prop:convex} allow us to deduce that (the entire) function $f$ is convex lower-semicontinuous. In a second step, based on the assumptions made on $g_{\mathsf{Q}}$ and the form of the function $\widetilde{g}_{\vartheta}$, we show that $f$ is also proper (\textit{i.e.,} its domain is non-empty), which completes the proof.\\

\subsubsection{Proof of Proposition~\ref{prop:minim}}
The proof uses three main steps. First, we show the existence of a unique infimum $\widehat{\vm}$ based on strict convexity and coercivity of $f$ with respect to $\vm$ (only). Then, using \cite[Proposition 8.35]{Livre1}, we restrict the analysis to a new function composed of the marginal function and remaining parts of $f$ independent of $\vm$. Writing the domain of this new function explicitly, we show that the necessary properties for \cite[Proposition 9.33]{Livre1} are satisfied; thus, it is proper convex lower-semicontinuous. Based on Assumption \ref{a:pminiii}, we finally show that this new function is coercive. Thus, a minimizer exists and is unique if strict convexity is achieved for $g_{\mathsf{Q}}$ and $\widetilde{g}_{\vartheta}$.
% %
% %--------------------------------------
% \subsection{Choice of the Regularizations}\label{sec:choice_of_reg}
% %--------------------------------------
% %
%
%--------------------------------------
\subsection{Regularization on $\vm$ and $\vQ$}\label{sec:choice_of_reg_Qm}
%(\NO{to be incorporated in previous subsection A?})
%--------------------------------------
%
As seen in Subsection~\ref{sec:study_cost_fct}, the assumptions on the regularizations $g_{\mathsf{m}}$ and $g_{\mathsf{Q}}$ are relatively straightforward to satisfy. In the following, we let $g_{\mathsf{m}} = \iota_{\{ \mathbf{0} \}}$ and 
%and focus on studying the regularizations on the variables $\mathbf{Q}$ and $\vthe$, for which prior information is more frequently available in practice.
%
% \subsubsection{Regularization on $\vQ$}
%----------------------------------------
consider a frequently encountered scenario where the precision matrix $\vC^{-1}$ is sparse. Note that the standard Graphical LASSO (GLASSO) problem~\cite{GLASSO08} can be obtained in the particular case where $\beta = 2$ by setting $g_{\mathsf{m}} = \iota_{\{\vzero\}}$, $g_{\vartheta}=\iota_{\{\boldsymbol{1}\}}$, and $g_{\mathsf{Q}}\colon \vQ \mapsto \lambda \|\vQ^2\|_{1}$ with $\lambda \in \RPP$. However, with the proposed re-parametrization, it is more direct to impose sparsity on the (new) variable $\vQ$ by choosing $g_{\mathsf{Q}}\colon \vQ \mapsto \lambda \|\vQ\|_{1}$. We will use the latter function for the experiments in Section~\ref{sec:4}. Another typical choice for enforcing sparsity is elastic net regularization where $g_{\mathsf{Q}} = \lambda \|\cdot\|_{1} + \frac{\epsilon}{2}\|\cdot\|_{\rm F}^2$ with $\lambda$ and $\epsilon \in \RPP$. It should be noted that elastic net regularization is strictly convex, and if $\widetilde{g}_{\vartheta}$ is also strictly convex, $f$ will admit a unique minimizer, as stated in Proposition~\ref{prop:minim}. 
\subsection{Regularization on $\vthe$}\label{sec:choice_of_reg_theta}
%----------------------------------------
\subsubsection{Choice of $g_{\vartheta}$}
%----------------------------------------
The choice of the regularization on $\vthe$ is central to our analysis. The goal is to ensure the convexity of the global cost function $f$ by compensating for the last term in \eqref{eq:newL}. The concavity of this term
will be counterbalanced by choosing a regularization leading to a coercive function.
%A way of counterbalancing this term is to combine the chosen regularization with a term weighted in a way that leads to a coercive function.} 
A simple choice for the regularization function $g_{\vartheta}$ allowing us to achieve this goal (while satisfying the requirements in Propositions~\ref{prop:convex} and \ref{prop:minim}) is the potential of a generalized Gamma distribution with scale parameter $\eta$, shape parameter $(\kappa+1)$, and exponent parameter $\alpha$, \textit{i.e.,}
\begin{equation}\label{e:potGgamma}
(\forall \vthe \in \RR^N)\quad
g_{\vartheta}(\vthe) =  \frac{1}{\eta^\alpha} \|\vthe\|_{\alpha}^\alpha +\kappa \sum_{n=1}^N \psi(\theta_{n})
\end{equation}
where $\psi$ has been defined in \eqref{eq:defpsi}, $\alpha \in [1,+\infty[$, $\eta \in \RPP$, and $\kappa \in ]K(1-1/\beta),+\infty[$.  %\NO{The choice of the latter parameters will not be further investigated in this work.} 
%--
We can then set
\begin{align*}
\widetilde{g}_{\vartheta,0}(\vthe) &= \frac{1}{\eta^\alpha} \|\vthe\|_{\alpha}^\alpha\\
\widetilde{g}_{\vartheta,1}(\vthe) &= \big(\kappa-K(1-1/\beta)\big) \sum_{n=1}^N \psi(\theta_{n}),
%\widetilde{g}_{\vartheta,1}(\vthe) &= \big(\kappa-K(1-1/\beta)-1\big) \sum_{n=1}^N \psi(\theta_{n}),
\end{align*}
which are functions satisfying Assumptions \ref{a:pminii} and \ref{a:pminiii} in Proposition \ref{prop:minim}. As emphasized above, coercivity is achieved by adjusting $\kappa$ such that the weight $\kappa-K(1-1/\beta)$ of the logarithmic term is positive. In addition, since $\widetilde{g}_{\vartheta,1}$ is then strictly convex, $\widetilde{g}_{\vartheta}$ is also strictly convex. It follows that the resulting cost function has a unique minimizer, provided that the chosen $g_{\mathsf{Q}}$ is also strictly convex.\\

\subsubsection{Impact on Cost Function $f$}
%---------------------------------------------
To gain better insight into the impact of regularization \eqref{e:potGgamma}, we study the average behaviour of the cost function $f$ with respect to $\vthe$ and the associated regularization parameters $\eta$, $\kappa$, and $\alpha$.\footnote{We will denote the true target value in this subsection by $\bar{\vthe}$ (resp. $\bar{\vtau}$) to distinguish it from the parameter or variable $\vthe$ (resp. $\vtau$).} Let us first isolate the terms of interest by introducing functions $(f_n)_{1\le n \le N}$ such that, for every $(\vQ,\vm,\vthe)\in \smp\times\RR^K\times \RPP^N$,
\begin{equation}
f(\vQ,\vm,\vthe) = \sum_{n=1}^N f_{n}(\theta_{n},\vQ,\vm)+
\Psi(\vQ) +g_{\mathsf{Q}}(\vQ)+g_{\mathsf{m}}(\vm),
\end{equation}
and for every $n \in \{1,\ldots,N\}$,
\begin{equation}
f_{n}(\theta_{n},\vQ,\vm) = \frac{\| \vQ \vy_{n}-\vm\|^\beta}{2\theta_{n}^{\beta-1}}
+ \frac{ \theta_{n}^\alpha}{\eta^\alpha} -\big(\kappa-K(1-1/\beta)\big) \log\theta_{n}.
\end{equation}
To focus our analysis on the variable $\vthe$, we make the simplifying assumption that $\vQ$ and $\vm$ have been perfectly identified with respect to the original statistical model. 
We can then restrict the analysis to the above function $f_n$ by looking at its average behaviour:
\begin{multline}\label{eq:expectation}
\mathsf{E}\{f_{n}(\theta_{n},\vQ,\vm)\} = \frac{\mathsf{E}\{\| \vQ \vy_{n}-\vm\|^\beta\}}{2\theta_{n}^{\beta-1}}
+ \frac{ \theta_{n}^\alpha}{\eta^\alpha}-\\ \big(\kappa-K(1-1/\beta)\big)  \log\theta_{n}.
\end{multline}
In \eqref{eq:expectation}, $(\vQ \vy_{n}-\vm)$ follows an MGGD$_K(\beta, \boldsymbol{0}, \overline{\tau}_n^2\,\ID)$, where $\overline{\tau}_n>0$ is the true value of the multiplicative perturbation factor in Model \eqref{e:statmodel} (with re-parametrizations \eqref{eq:reparam_Q}--\eqref{eq:reparam_theta}). We then have
\begin{equation}
\mathsf{E}\{\| \vQ \vy_{n}-\vm\|^\beta\} = \frac{2K}{\beta}\, \overline{\theta}_n^{\beta-1}
\end{equation}
with $\overline{\theta}_n 
= \overline{\tau}_n^{\beta/(\beta-1)}$,
and
\begin{equation}
\mathsf{E}\{f_{n}(\theta_{n},\vQ,\vm)\} = K(1-1/\beta) \overline{f}(\theta_{n}),
\end{equation}
 where
\begin{equation}\label{eq:regthetamean}
\overline{f}(\theta_{n})
= \frac{\overline{\theta}_n^{\beta-1}}{(\beta-1)\theta_{n}^{\beta-1}}
+ \frac{ \beta \theta_{n}^\alpha}{K(\beta-1)\eta^\alpha} -\big(\overline{\kappa}-1\big)  \log\theta_{n}
\end{equation}
and 
\begin{equation}
    \overline{\kappa} = \kappa\beta/(K(\beta-1)).
\end{equation}
With the conditions imposed on $\kappa$, one has $\overline{\kappa} > 1$.
The derivative of $\overline{f}$ at $\theta_{n}$ is
\begin{equation}
\overline{f}'(\theta_{n})= \frac{1}{\theta_{n}}\left(1-
\Big(\frac{\overline{\theta}_n}{\theta_n}\Big)^{\beta-1}
+ \frac{ \alpha\beta \theta_{n}^{\alpha}}{K(\beta-1)\eta^\alpha} -\overline{\kappa}\right).
\end{equation}
Moreover, one has $\overline{\theta}_n = 1$ for unperturbed data. In this case, if the scale parameter 
\begin{equation}\label{e:etaopt}
\eta = \Big(\frac{\alpha\beta}{K(\beta-1)\overline{\kappa}}\Big)^{1/\alpha} \text{ or equivalently, } \eta = \Big(\frac{\alpha}{\kappa}\Big)^{1/\alpha},
\end{equation}
then $\overline{f}$ is decreasing over $]0,1]$ and increasing over $[1,+\infty[$ (see Appendix~\ref{a:studyfbar}). As expected, the minimum of this function is thus attained at $\widehat{\theta}_n = 1$. 
%\NO{Does it make sense to also study the behaviour of fbar with respect to eta?}
%
For a given value of $\overline{\kappa}$, the choice for $\eta$ in \eqref{e:etaopt} makes function $\overline{f}$ independent of the vector dimension $K$, and we have, in this case, 
\begin{equation}\label{e:ovfopteta}
\overline{f}(\theta_{n})
= \frac{\overline{\theta}_n^{\beta-1}}{(\beta-1)\theta_{n}^{\beta-1}}
+ \frac{ \overline{\kappa}\theta_{n}^\alpha}{\alpha} -\big(\overline{\kappa}-1\big)  \log\theta_{n}.
\end{equation}

In Appendix \ref{a:studyfbar}, we show that the above function has a unique minimizer $\widehat{\theta}_n$. If $\overline{\theta}_n > 1$, $\widehat{\theta}_n \in ]1,\overline{\theta}_n[$, which means that a bias is introduced by the regularization. As $\widehat{\theta}_n$ is an increasing function of $\overline{\theta}_n$, the method remains effective in reducing the influence of outliers. In addition, when $\overline{\theta}_n > 1$, $\widehat{\theta}_n$ is a decaying function of $\alpha$ (resp. $\overline{\kappa}$). In other words, a stronger regularization comes at the expense of a larger bias in estimating the perturbation parameter $\tau_n$. This suggests choosing both $\alpha$ and $\overline{\kappa}$ close to 1.\\ 
%(\NO{trade-off in practice?})%(\NO{trade-off no-regularization/bias instead of lowest possible parameters? there must be some circumstances in favour of introducing a larger amount of bias})

\begin{figure}
    \centering
    \includegraphics[width=\linewidth]{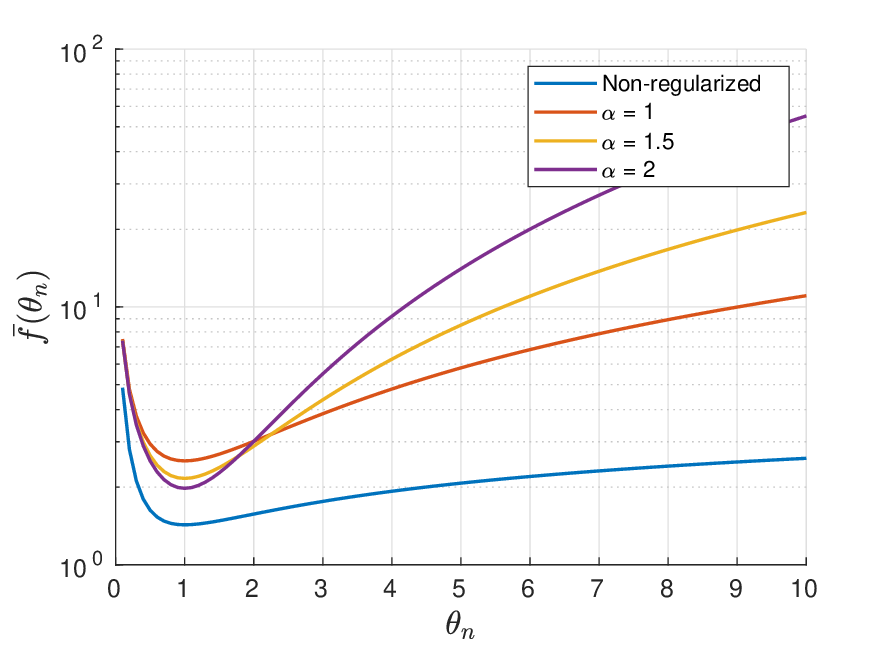}%{plot-reg.eps}
    \caption{Graph of function $\overline{f}$ in \eqref{eq:regthetamean} when $\eta$ is given by \eqref{e:etaopt}, $\beta = 1.7$, and $\overline{\theta}_n = 1$. In blue, the non-regularized case when $\eta \to \pinf$ and $\kappa = 0$. Other plots correspond to $\overline{\kappa} = 1.1$ with increasing values of $\alpha$, \textit{i.e.,} $\alpha = 1$ in red, $\alpha = 1.5$ in yellow, and $\alpha = 2$ in purple. The ordinate axis is in the log scale.} 
    \label{fig:plot-reg}
\end{figure}

\begin{figure}
    \centering
    \includegraphics[width=\linewidth]{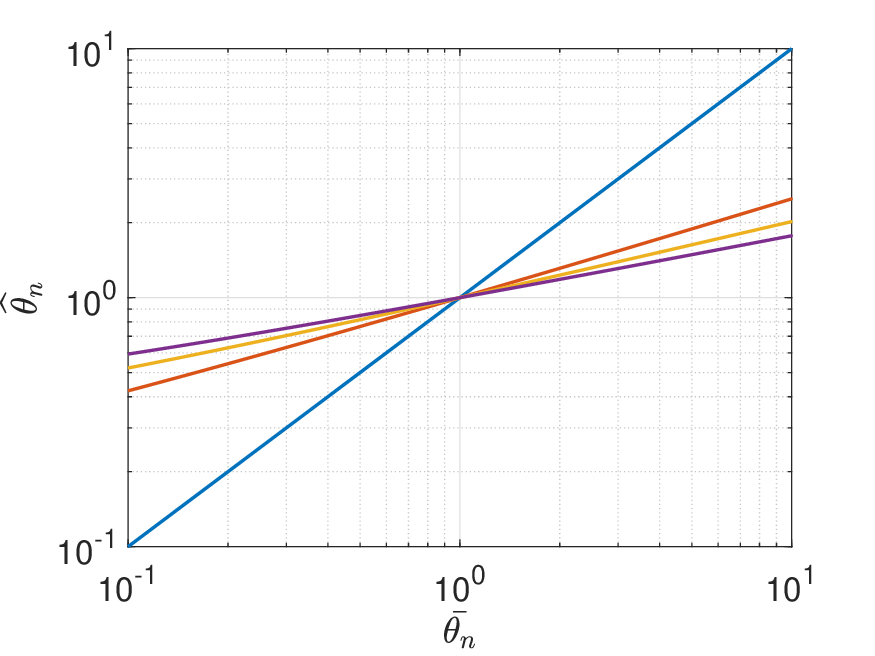}%{plot-opt-theta.eps}
    \caption{Variations of $\widehat{\theta}_n$ versus $\overline{\theta}_n$ based on a numerical solution of \eqref{e:opttheta} when $\beta = 1.7$. In blue, the non-regularized case where $\kappa = 0$. Other plots correspond to $\overline{\kappa} = 1.1$ with increasing values of $\alpha$, \textit{i.e.,} $\alpha = 1$ in red, $\alpha = 1.5$ in yellow, and $\alpha = 2$ in purple. Both axes are in the log scale.}
    \label{fig:plot-opt-theta}
\end{figure}

Plots of the graph of the function $\overline{f}$ for different values of $\alpha$ are shown in Figure~\ref{fig:plot-reg}. As expected, when $\alpha$ increases, stronger perturbations $\theta_n$ are penalized more heavily. Figure~\ref{fig:plot-opt-theta} shows the variations of $\widehat{\theta}_n$ as a function of $\overline{\theta}_n$ for the same parameter values. The introduced bias can be seen in the increasing gap between the target $\overline{\theta}_n$ and $\widehat{\theta}_n$ as $\alpha$ increases. We can also notice that the bias increases as the perturbations become larger for a given value of $\alpha$. Note that although a larger value of $\alpha$ introduces more bias, it can also benefit the convergence speed of optimization algorithms. For example, when $\alpha = 2$, $\overline{f}$ is a strongly convex function for which we can typically expect a linear convergence rate of optimization algorithms. This suggests that a trade-off between acceptable bias and reasonable computational speed is to be considered in practice.
%
%This means that adding a regularization on $\vthe$ comes at the cost of robustness against the perturbations. Notice that the unregularized case (blue curve in Fig.~\ref{fig:plot-reg}) is the most robust, \textit{i.e.,} less weight is put on the cost function when $\theta_n$ becomes strong. Similarly, setting $\alpha=1$ is the most robust choice in the regularized case. In practice, a trade-off between robustness (using lower values of $\alpha$) and a more efficient minimization (increasing the value of $\alpha$) should be considered. \textit{\NO{(NO: translate these findings into practical choices for alpha + explain why minimization is faster, curvature?)}}\\ 
%
%
%\end{remark}
%\nonumber\\
%& \ge \Psi(\vQ)+g_{\mathsf{Q}}(\vQ)+\widetilde{g}_{\vartheta}(\vthe)+g_{\mathsf{d}}(\vd)
%\end{align}
%
% \textit{\NO{It is worth noticing that most existing works focus on Gaussian models.} We will now focus on how to minimize this function.\NO{To preserve convexity, some assumptions on the regularizations are, however, necessary. It can be shown that the regularizations in \eqref{missing} preserve the convexity of the f. The proofs will be made available in a forthcoming paper for lack of space.}}
%
% With the choice of the regularization suggested above, the resulting cost function is convex, and it has a unique minimizer under mild conditions.
% Due to the lack of space, formal proof will be detailed in a forthcoming paper. This function can be minimised by proximal primal-dual algorithms for which the proximity operators of all components of the function are available in closed form~\cite{Livre1,Com18,Combettes2011}. Convergence is guaranteed \cite{Komodakis2015,Combettes2021} and reached in a few iterations in our experiments (see Section~\ref{sec:4}).
%
%\clearpage
%----------------------------------
\section{Proximal Primal-Dual Algorithm}
\label{se:PFalgo}
%----------------------------------
Minimizing cost function \eqref{e:costfunc} requires handling various regularizations. Proximal algorithms can efficiently tackle such minimization problems: regularization terms are dealt with using proximity operators, usually in closed form~\cite{Livre1,Com18,Combettes2011}. In this section, we will show that Problem \eqref{e:costfunc} can be reformulated as the minimization of a sum of compositions of convex functions and linear operators, which can be solved using a proximal primal-dual algorithm.
The primal space is here $\mathcal{H} = \mathcal{S}_K\times  \RR^K\times \RR^N$, endowed with the norm:
\begin{equation}
(\forall \mathbf{p} = (\vQ,\vm,\vthe) \in \mathcal{H})
\;\;
\|\mathbf{p}\|_{\mathcal{H}}
= \sqrt{\|\vQ\|_{\rm F}^2+\|\vm\|^2+\|\vthe\|^2}.
\end{equation}
The dual space is $\mathcal{G}_1\times 
\mathcal{G}_2$ with
$\mathcal{G}_1 = (\RR^K)^N\times \RR^N$
and $\mathcal{G}_2 = \mathcal{S}_K\times \RR^N$. The first space is endowed with the norm
\begin{multline}\label{eq:w1}
\big(\forall \mathbf{v}_1= \big((\mathbf{u}_n)_{1\le n \le N},
\vthe)\in \mathcal{G}_1\big)\\
\|\mathbf{v}_1\|_{\mathcal{G}_1}
= \sqrt{\sum_{n=1}^N\|\mathbf{u}_n\|^2
+ \omega_1 \|\vthe\|^2},
\end{multline}
while the second one is equipped with the norm
\begin{equation}\label{eq:w2}
\big(\forall \mathbf{v}_2= \big(\vQ,
\vthe)\in \mathcal{G}_2\big)\;\;
\|\mathbf{v}_2\|_{\mathcal{G}_2}
= \sqrt{\|\vQ\|_{\rm F}^2
+ \omega_2\|\vthe\|^2},
\end{equation}
where $(\omega_1,\omega_2) \in \RPP^2$. In the proposed algorithm, the latter parameters make the update of the $\vthe$ estimates faster by introducing more flexibility. Note that all the described product spaces are finite-dimensional Hilbert spaces.

We reformulate the optimization problem as 
\begin{equation}\label{e:pboptimPD}
    \minimize{\mathbf{p}\in \mathcal{H}} F(\mathbf{p})
    + G_1(\mathbf{L}_1\mathbf{p})
    +G_2(\mathbf{L}_2\mathbf{p}),
\end{equation}
where $\mathbf{L}_1$ and $\mathbf{L}_2$ are linear operators such that
\begin{align}
\mathbf{L}_1 &\colon  \mathcal{H} \to
\mathcal{G}_1\colon
(\vQ,\vm,\vthe) \mapsto (\mathcal{T}(\vQ,\vm),\vthe)\nonumber\\
\mathbf{L}_2&\colon \mathcal{H}\to \mathcal{G}_2\colon
(\vQ,\vm,\vthe) \mapsto (\vQ,\vthe),
\end{align}
and $\mathcal{T}$ is given by \eqref{e:defcalT}. 
With the choice of the regularisation functions made
in Sections \ref{sec:choice_of_reg_Qm} and 
and \ref{sec:choice_of_reg_theta},
%\FP{and under the assumptions of Proposition~\ref{prop:minim}}, 
$F$ and $G_2$ are convex functions such that
\begin{align}
F(\vQ,\vm,\vthe)
&= \Psi(\vQ)+g_{\mathsf{m}}(\vm)+
\widetilde{g}_{\vartheta,1}(\vthe)\nonumber\\
G_2(\vQ,\vthe) & = 
g_{\mathsf{Q}}(\vQ)+\widetilde{g}_{\vartheta,0}(\vthe),
\end{align}
and $G_1=\Phi$ is given by \eqref{e:defPhi}. 
In addition, the existence of a solution to the minimization problem follows from Proposition~\ref{prop:minim}.
Since $F$ and $G_2$ are separable functions, their proximity operators can be calculated componentwise.

To solve \eqref{e:pboptimPD}, a standard solution is the Chambolle-Pock algorithm~\cite{Chambolle2011}, which reads
\begin{align}\label{e:CPbasic}
&\mbox{For $k=0,1,\ldots$}\nonumber\\
&\left\lfloor
\begin{array}{l}
\mathbf{p}_{k+1} = \prox_{\gamma F}(\mathbf{p}_{k}-\gamma (\zeta_1 \mathbf{L}_1^* \mathbf{v}_{1,k}+\zeta_2 \mathbf{L}_2^*
\mathbf{v}_{2,k}))
\\
\widetilde{\mathbf{p}}_k = 2\mathbf{p}_{k+1}-\mathbf{p}_{k}\\
\mathbf{v}_{1,k+1}=
\Big(\mathbf{Id}-\prox_{\zeta_1^{-1}G_1}\Big)
(\mathbf{v}_{1,k}+\mathbf{L}_1 \widetilde{\mathbf{p}}_k)
\\
\mathbf{v}_{2,k+1}=
\Big(\mathbf{Id}-\prox_{\zeta_2^{-1}G_2}\Big)
(\mathbf{v}_{2,k}+\mathbf{L}_2 \widetilde{\mathbf{p}}_k),
\end{array}
\right. 
\end{align}
starting from initial values
$\mathbf{p}_{0}\in \mathcal{H}$, $\mathbf{v}_{1,0}\in \mathcal{G}_1$,
and $\mathbf{v}_{2,0}\in \mathcal{G}_2$. The variables $(\mathbf{v}_{i,k})_{k\ge 0}$ with $i\in \{1,2\}$ are dual variables.
Hereabove $\mathbf{L}_1^*$ (resp. 
$\mathbf{L}_2^*$) is the adjoint of 
$\mathbf{L}_1$ (resp. $\mathbf{L}_2$),
and $(\gamma,\zeta_1,\zeta_2)$ are positive scalar parameters. To ensure the algorithm convergence, these parameters must be chosen such that 
\begin{equation}\label{e:condconv}
\gamma(\zeta_1 \|\mathbf{L}_1\|^2_{\rm S}
+ \zeta_2 \|\mathbf{L}_2\|^2_{\rm S}) < 1,
\end{equation}
where $\|\cdot\|_{\rm S}$ is the operator norm. From the calculations of the norms of $\mathbf{L}_1$ and $\mathbf{L}_2$ in Appendix \ref{a:propL1L2}, we deduce that a sufficient condition for \eqref{e:condconv} to be satisfied is
\begin{equation}\label{e:condconvf}
\zeta_1 
\max\{\|\mathbf{Y}\|_{\rm S}^2,\omega_1\}
+ \zeta_2 \max\{1,\omega_2\} < \gamma^{-1},
\end{equation}
where
\begin{align}\label{e:defY}
\mathbf{Y} 
&= \begin{bmatrix}
\vy_1 & \ldots & \vy_N\\
1 & \ldots & 1
%\mathbf{Y}\\
%-\mathbf{1}_N^\top
\end{bmatrix}.
%\nonumber\\
%\mathbf{Y} 
%&= \begin{bmatrix}
%\vy_1 & \ldots \vy_N
%\end{bmatrix},
\end{align}
%and $\mathbf{1}_N=[1,\ldots,1]^\top\in \RR^N$
%On the other hand, it follows from the structure
%of $\mathbf{Z}$ that

We can then rewrite Algorithm \eqref{e:CPbasic} to make the role of each involved variable more explicit. Let us set, at each iteration number $k$, 
$\mathbf{p}_k= (\vQ_k,\vm_k,\vthe_k)$,
$\widetilde{\mathbf{p}}_k= (\widetilde{\vQ}_k,\widetilde{\vm}_k,\widetilde{\vthe}_k)$,
$\mathbf{v}_{1,k} = ((\mathbf{u}_{k,n})_{1\le n \le N},\vthe_{1,k}^\#)$,
and $\mathbf{v}_{2,k} =
(\vQ_k^\#,\vthe_{2,k}^\#)$, and let us define
$\zeta_3=\omega_1 \zeta_1$ and $\zeta_4 = \omega_2 \zeta_2$. $\vthe_{1,k}^\#$ (resp. $\vthe_{2,k}^\#$) corresponds to the dual variable of $\vthe_{k}$ in the space $\mathcal{G}_1$ (resp. $\mathcal{G}_2$), while $\vQ_k^\#$ stands for the dual variable of $\vQ_k$ in the space $\mathcal{G}_2$. From the expressions of the adjoint operators of $\mathbf{L}_1$ and $\mathbf{L}_2$ derived in Appendix \ref{a:propL1L2}, we obtain the following iterative algorithm:
%\begin{align}
\begin{equation}
\begin{array}{l}
%&
\mbox{For $k=0,1,\ldots$}\nonumber\\
%&
\left\lfloor
\begin{array}{l}
\displaystyle \widehat{\vQ}_k 
= \vQ_k-\gamma\Big(\zeta_1\sum_{n=1}^N \mathbf{u}_{k,n} \vy_n^\top
+\zeta_2 \vQ_k^\#\Big)\\
\vQ_{k+1} = \prox_{\gamma \Psi}
\big((\widehat{\vQ}_k+\widehat{\vQ}_k^\top)/2\big)\\
\displaystyle\vm_{k+1} = \prox_{\gamma g_{\mathsf{m}}}
\Big(\vm_k+\gamma \zeta_1 \sum_{n=1}^N \mathbf{u}_{k,n}\Big)\\
\vthe_{k+1} = \prox_{\gamma \widetilde{g}_{\vartheta,1}}\big(\vthe_k-
\gamma (\zeta_3 \vthe_{1,k}^\#+
\zeta_4 \vthe_{2,k}^\#)\big)\\
\widetilde{\vQ}_k = 2 \vQ_{k+1}-\vQ_k\\
\widetilde{\vm}_k = 2 \vm_{k+1}-\vm_k\\
\widetilde{\vthe}_k = 2 \vthe_{k+1}-\vthe_k\\
\mbox{For $n=1,\ldots,N$}\\
\left\lfloor
\begin{array}{l}
(\mathbf{u}_{k+1,n},\theta_{1,k+1,n}^\#)\\
\;\;= \Big(\mathbf{Id}-\prox_{\varphi}^{\zeta_1^{-1}, \zeta_3^{-1}}\Big)
(\mathbf{u}_{k,n}+\widetilde{\vQ}_k \vy_n-\widetilde{\vm}_k,
\theta_{1,k,n}^\#+\widetilde{\theta}_{k,n})
\end{array}
\right.\\
\vQ_{k+1}^\# =
\Big(\mathbf{Id}-\prox_{\zeta_2^{-1}g_{\mathsf{Q}}}\Big)
(\vQ_k^\#+\widetilde{\vQ}_k)\\
\vthe_{2,k+1}^\# = 
\Big(\mathbf{Id}-
\prox_{\zeta_4^{-1}\widetilde{g}_{\vartheta,0}}\Big)(\vthe_{2,k}^\#+\widetilde{\vthe}_k).
\end{array}
\right. 
\end{array}
\end{equation}\\

The required proximity operators are provided in Appendix~\ref{Apx:prox}. 
%
%\NO{Convergence is guaranteed~\cite{Komodakis2015,Combettes2021} and reached in a few iterations in our experiments (see Section~\ref{sec:4})}.
%
% \subsection{Proximal Primal-Dual Algorithm}
%
%---------------------------------------
\section{Experiments}\label{sec:4}
%---------------------------------------
%
This section evaluates the proposed method in comparison with various state-of-the-art approaches. For the proposed method, the experiments are carried out by imposing sparsity on the precision matrix and regularization \eqref{e:potGgamma} on the perturbations (see Subsections~\ref{sec:choice_of_reg_Qm} and \ref{sec:choice_of_reg_theta}). We present results with varying $N$ (sample size) and $K$ (dimension) values. In particular, we study the standard case $N>K$ and the high-dimensional setting where $N\ll K$. The impact of sparse regularization on the precision matrix is investigated by considering dense and sparse matrices. Finally, we study the behaviour of the estimators with varying perturbation levels.\\ 
% We particularly highlight the benefits of sparse regularization on the precision matrix. For the real data set, we propose to study the clustering performance of different MGGD parameter estimation methods and compare them to classical clustering algorithms such as K-means, DBSCAN and HDBSCAN, which are particularly suited to data with outliers.
% % 
% \NO{We also consider two simulation scenarios with different precision matrix types and data sizes. (mention dense and sparse?)} We compare the proposed method to the robust Tyler's estimates and the empirical statistics in both the observed (\textit{i.e.,} perturbed) and ideal non-perturbed cases. \NO{Note that a fixed-point update is used to estimate the mean in Tyler's method. Although this approach has no convergence guarantees, it is frequently used in practice~\cite{missing}}.} 
% This section presents simulation results with varying $N$ values.

Our comparisons aim to evaluate the estimators' consistency and mean square error (MSE) with respect to the true parameters. The consistency of an estimated matrix $\bb{\hat{A}}$ is quantified by $\| \bb{\hat{A}} - \bb A \|_{\rm F}$,  whereas the MSE is $\mathsf{E} \left(\| \bb{\hat{A}} - \bb A \|^2_{\rm F}\right)$, where $\bb A$ is the true matrix. In our experiments,
the expectation is computed empirically by averaging $N_{\textrm{MC}} = 10^4$ Monte Carlo runs except for the high-dimensional case where we use $N_{\textrm{MC}} = 10^3$. 

%Computing MSEs requires averaging errors over $I$ Monte Carlo runs for a given experiment: the MSE of an estimated matrix $\bb{\hat{A}}$ is thus computed as $\widehat{MSE}_n = \frac{1}{I K^2}\sum\limits_{i=1}^I\|\bb{\hat{A}}_i - \bb A \|^2_{\rm F}$, for $n=1, \hdots, N_{MC}$. We choose $I=100$ for all experiments. Then, the final MSE is computed from the $N_{MC}/I$ estimated $\widehat{MSE}_n$'s.
%
\subsection{State-Of-The-Art Methods}
In the standard case when $N>K$, the proposed method is compared to the empirical statistics and robust Tyler's estimates in both perturbed and ideal non-perturbed ($(\forall n)$ $\tau_n = 1$) scenarios. Note that a fixed-point update estimates the mean in Tyler's method. Although this approach has no convergence guarantees, it is frequently used in practice~\cite{frontera2015hyperspectral}. In the high-dimensional setting ($N\ll K$), where Tyler's estimates do not exist, we show the results of the regularized Tyler's method~\cite{pascal2014generalized,ollila2014regularized}. In addition, we compare the proposed method to classical GLASSO~\cite{GLASSO08} (see Subsection~\ref{sec:choice_of_reg_Qm}) and its robust extension proposed in~\cite{Ollerer2015}. Note that GLASSO and its robust version also use a sparse regularization on the precision matrix. In~\cite{Ollerer2015}, several robust GLASSO extensions are proposed by replacing the classical GLASSO input with a robust covariance matrix estimate. Among the estimators presented in~\cite{Ollerer2015}, we investigate GlassoGaussQn (the best-performing one), which computes the robust covariance matrix using the Qn-estimator of scale and a robust correlation matrix.

% The comparison uses the estimated consistency concerning the true parameters. More precisely, the estimated consistency of $\bb{\hat{A}}$ is defined as $\| \bb{\hat{A}} - \bb A \|_{\rm F}$, where $\bb A$ is the true parameter. Finally, all experiments are performed by averaging $1000$ Monte Carlo runs. 
%
%----------------------------------
% \subsection{Experiments on Synthetic Data}
%----------------------------------
%
%----------------------------------
\subsection{Simulation Parameters}
%----------------------------------
%
Unless indicated otherwise, $\beta= 1.5$, the maximum perturbation level is $\tau_{\text{max}} = 5$, and the proportion of corrupted data is set to $p_\tau = 0.3$ (when invariant). The mean vector is drawn from the standard normal distribution in all experiments. The parameters of the gamma prior are set to $\kappa= 1.1\, K(1-1/\beta)$ and $\alpha = 1$, which are parameters satisfying the conditions in Subsection~\ref{sec:choice_of_reg_theta}. The sparse regularization parameter $\lambda$ is a user-defined parameter; in this work, it is adjusted automatically by providing the desired sparsity level (\textit{i.e.,} the proportion of zero entries in the generated matrices). %$\vtau$ is generated with random perturbations with probability $p=0.1\%$
%
%----------------------------------
\subsection{Experiments with $N>K$}
%----------------------------------
%
The first experiments study two different cases of \textit{i.i.d.} $K=20$-dimensional data vectors $(\vx_{n})_{1\le n\le N}$ distributed first according to an MGGD$_K(\beta, \mathbf 0, \vC_1)$ where $\vC_1^{-1}$ is a \textit{sparse} precision matrix, and then, an MGGD$_K(\beta,\mathbf 0,\vC_2)$ with a \textit{dense} precision matrix $\vC_2^{-1}$.\footnote{The observed samples $(\vy_n)_{1\le n\le N}$ follow an MGGD$_K(\beta, \boldsymbol{\mu}, \tau_n^2\,\vC_1)$ (or an MGGD$_K(\beta, \boldsymbol{\mu}, \tau_n^2\,\vC_2)$) according to \eqref{e:statmodel}.} The precision matrix $\vC_1^{-1}$ is modelled by an auto-regressive (AR) process of order $3$ such that
\begin{equation}
\vC_1^{-1}(i,j) =
\begin{cases}
       \rho^{| i-j |}, &\!\!\!\textrm{ for }\ | i\!-\!j | = 0 \textrm{ and } i\in \{1,\hdots, K\} \\
       %\rho^{| i-j |}, 
       &\!\!\!\textrm{ for }\ | i\!-\!j | = 1 \textrm{ and } i\in \{2,\hdots,\!K\!-\!1\}\\
       %\rho^{| i-j |}, 
       & \!\!\!\textrm{ for }\ | i\!-\!j | = 2 \textrm{ and } i\in \{3,\hdots,\!K\!-\!2\}\\
       0 &\!\!\!\text{ otherwise.}
    \end{cases} 
\end{equation} 
resulting in a tri-diagonal sparse matrix. The entries of the dense matrix $\vC_2^{-1}$ are given by
\begin{equation}
(\forall (i,j) \in \{1,\hdots, K\}^2),\quad 
\vC_2 ^{-1}(i,j) = \rho^{| i-j |},
\end{equation}
\noindent where $\rho = 0.5$ is the correlation coefficient used to generate both precision matrices.
%
% %
% \begin{figure*}[!htb]
% \begin{minipage}[t]{\linewidth}%
% \centering
% \includegraphics[width=\textwidth]{FINAL AR MSE.eps}
% \centering (a)
% \end{minipage}
% \begin{minipage}[t]{\linewidth}%
% \centering
% \includegraphics[width=\textwidth]{FINAL AR CONSISTENCY.eps}
% \centering (b)
% \end{minipage}
% \hfill
% \caption{(a) MSEs and (b) consistencies of different mean, covariance and precision matrix estimators for the experiments with a sparse AR precision matrix.}
% \label{fig:exp1_AR}
% \end{figure*}
% %
%
\begin{figure*}[!htb]
\centering
\includegraphics[width=1.05\textwidth]{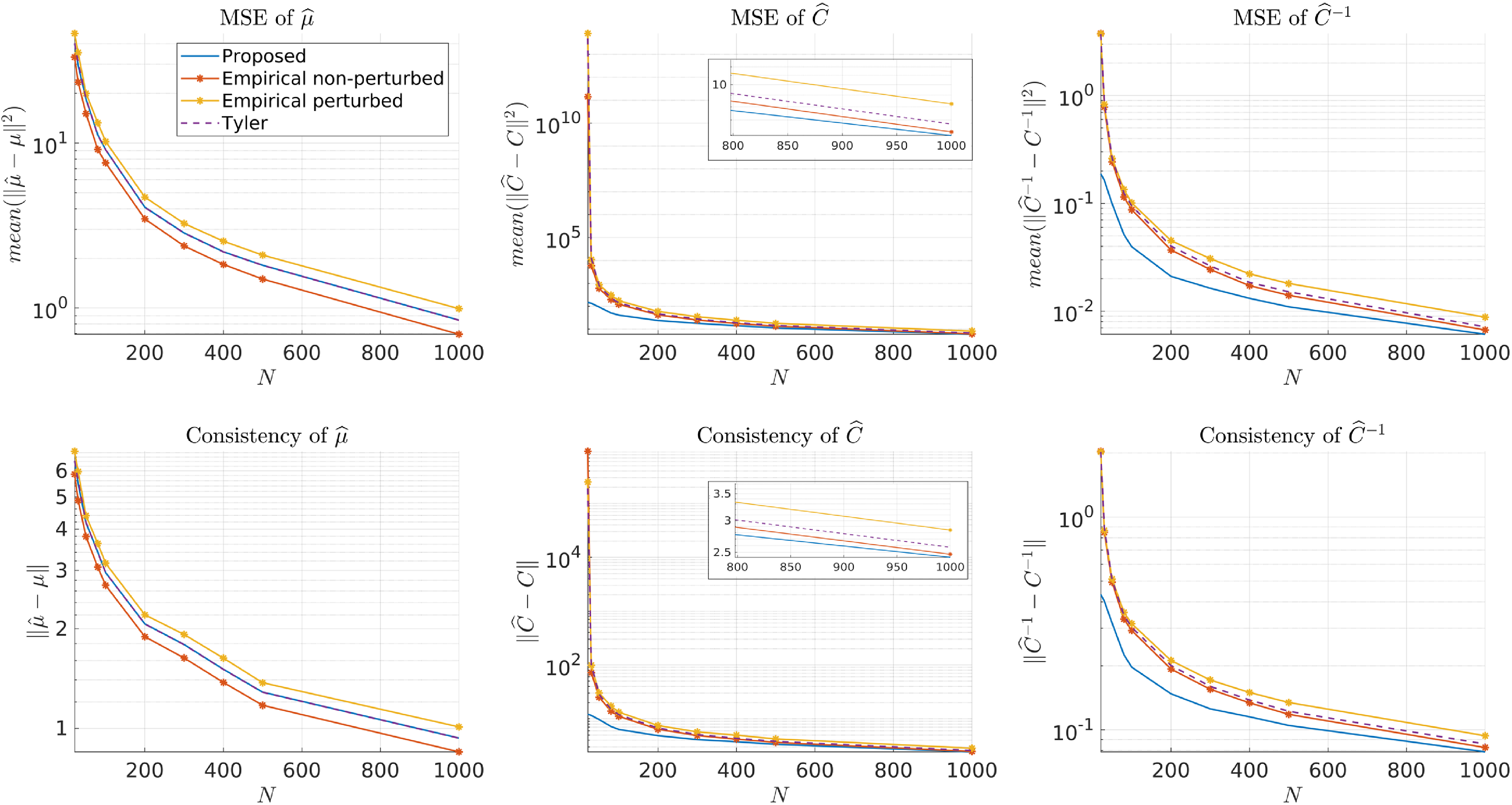}
\caption{MSEs and consistencies of different mean, covariance and precision matrix estimators for the experiments with a sparse AR precision matrix.}
\label{fig:exp1_AR}
\end{figure*}
Fig.~\ref{fig:exp1_AR} shows the evolution of the MSE and consistency of the obtained mean, covariance, and precision matrices for different values of $N$ in the case of a sparse precision matrix. As expected, the best mean estimates correspond to the empirical ones in the ideal case. The least accurate/consistent are those obtained with the empirical estimator in the perturbed case. Comparing the proposed method and Tyler's estimates reveals an almost identical performance. This is interesting because, unlike Tyler's method, the proposed method guarantees convergence in the unknown mean case.
On the other hand, the results obtained by the proposed method for covariance and precision matrix estimation show a notable improvement over all other approaches. These improvements in MSE and consistency are even more noticeable for small sample sizes, indicating that the proposed method performs relatively well with a limited number of observations. The superior performance of the proposed method can be explained by the direct estimation of the precision matrix, strengthened by an appropriate sparse regularization.
\begin{table}[!b]
\centering
\renewcommand*{\arraystretch}{1.3}
\setlength\tabcolsep{1.5pt}
%\caption{MSEs of the dense precision matrices estimated using the empirical perturbed, Tyler's, and proposed methods for increasing values of $N$.}
\caption{MSEs obtained for the experiments with a dense precision matrix for increasing sample sizes $N$.}
\resizebox{\linewidth}{!}{%
\begin{tabular}{|c|c |c |c || c| c| c || c |c |c |} 
\hline
   \multirow{2}{*}{$N$} & \multicolumn{3}{c ||}{Empirical} &\multicolumn{3}{c||}{Tyler} &\multicolumn{3}{c|}{Proposed}\\ 
 %  \midrule
  \cline{2-10}
 & $\mu$ & $\vC_2$  &  $\vC_2^{-1}$ & $\mu$  & $\vC_2$   &$\vC_2^{-1}$&$\mu$  & $\vC_2$   & $\vC_2^{-1}$ \\
 \hline
21 & 42.51  &  1.29e12 & 3.925  & \textbf{37.94}  & 1.10e12  & 3.950 & 40.54&\textbf{1.90e4}  & \textbf{0.629} \\
30 & 33.23 &  7.03e3 &  0.732 & \textbf{29.37} &  8.05e3  &0.791 & 29.98& \textbf{1.65e3} &\textbf{0.330}\\
50 &  16.73 & 766.25   & 0.262  & \textbf{14.75} & 651.59   & 0.265& 14.85& \textbf{419.14}  & \textbf{0.191} \\
 80 & 11.43 & 235.91  & 0.127  &\textbf{9.99} &   177.13 &  0.118 &10.04 & \textbf{147.80}  & \textbf{0.100} \\
 100 & 8.74 &  145.11 &  0.100& \textbf{7.77} & 119.10   &  0.092 & 7.80&  \textbf{99.53} & \textbf{0.079}\\
 200 & 4.40 & 47.14  & 0.045  &  \textbf{4.07}& 38.55   & 0.040  & 4.06&  \textbf{35.51}& \textbf{0.037} \\
 300 &  2.99& 27.85  & 0.029 & \textbf{2.59}&  22.31  & 0.025  & 2.60& \textbf{21.50} & \textbf{0.024} \\
 400 & 2.23 & 20.14  & 0.022 & \textbf{1.98} &  15.49  &  0.019& \textbf{1.98}&  \textbf{15.25} & \textbf{0.018 }\\
 500 & 1.93 & 15.00  & 0.018 &\textbf{1.66} &  12.00  & \textbf{0.015}& 1.67&  \textbf{ 11.8}5&  \textbf{0.015}\\
 1000 & 0.97 & 6.89  &  0.009& \textbf{0.84} &  5.43  & \textbf{0.007} &\textbf{0.84}&  \textbf{5.42} &\textbf{0.007} \\
 \hline
\end{tabular}%
}
\label{tab:dense1}
\end{table}
The benefits of sparse regularization in this first experimental scenario can be understood by studying the performance of dense precision matrix estimation (Table~\ref{tab:dense1}). Here, the MSE gap between different methods is less pronounced, probably due to the minor impact of sparse regularization. However, it is worth noting that the performance of the proposed method remains superior to the other approaches for covariance and precision matrix estimation. As in the sparse case, the MSEs of the mean are similar for the proposed and Tyler's methods. 
% The much better results for the precision matrix can be explained by the fact that the proposed method estimates this quantity directly rather than performing a matrix inversion of the estimated covariance matrices (which may result in numerical issues). The convergence of the total cost is shown in Fig.~\ref{fig:exp12_cvg}-(a). This figure shows the average evolution of the cost function over all Monte Carlo experiments.
%
%\NO{EXECUTION TIME COMPARISON, meme crit perf, meme nbr iterations or something. Should be faster than fixed-point a cause d' inversion de matrice ?}

\begin{figure}[!t]
\begin{minipage}[t]{.49\linewidth}%
\centering
\includegraphics[width=\textwidth]%{tauexp_averaged_MEP2.eps}\\
{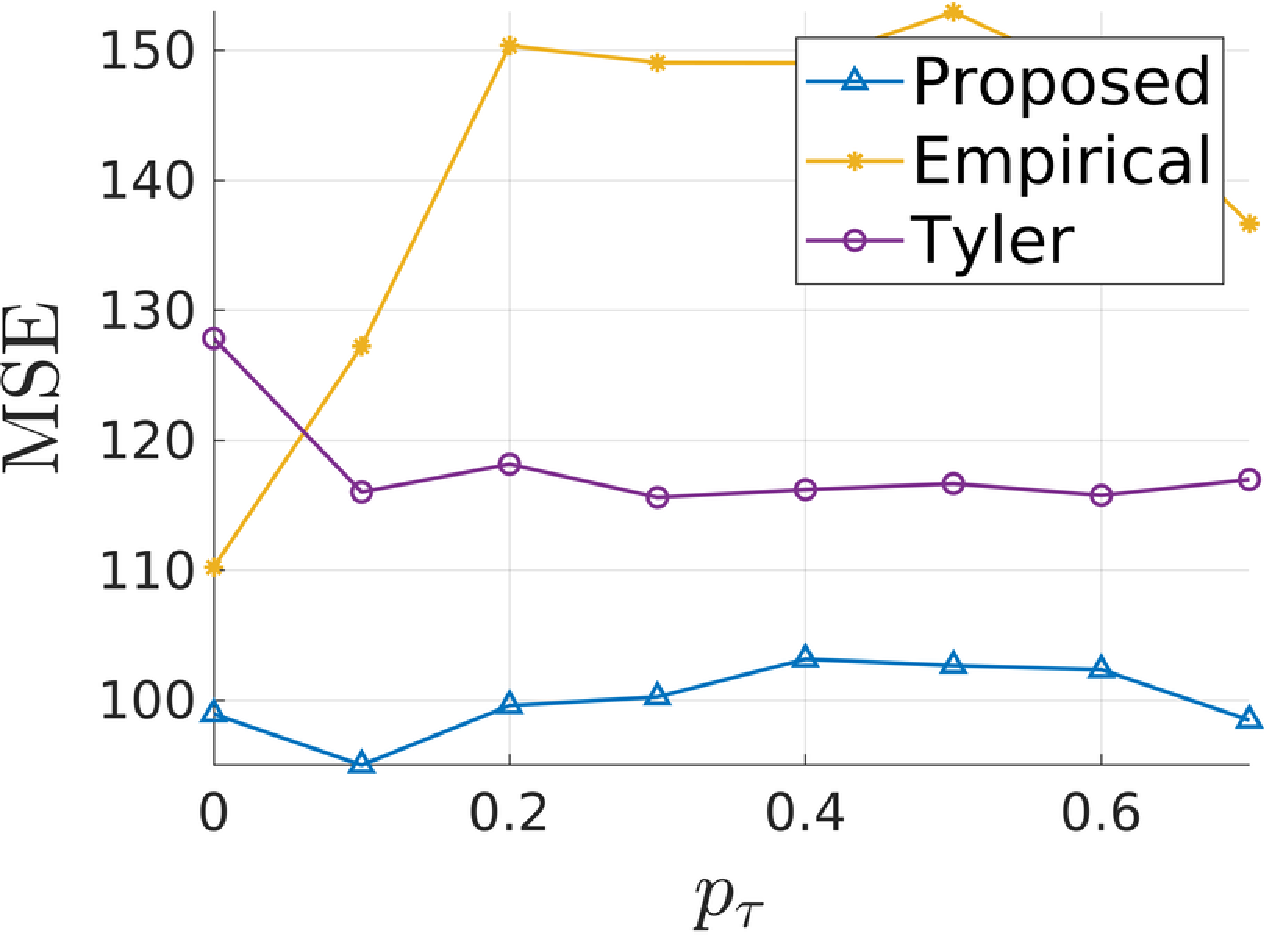}
\centering (a)
\end{minipage}
\begin{minipage}[t]{.49\linewidth}%
\centering
\includegraphics[width=\textwidth]%{tauexp_averaged_gauss2.eps}\\
{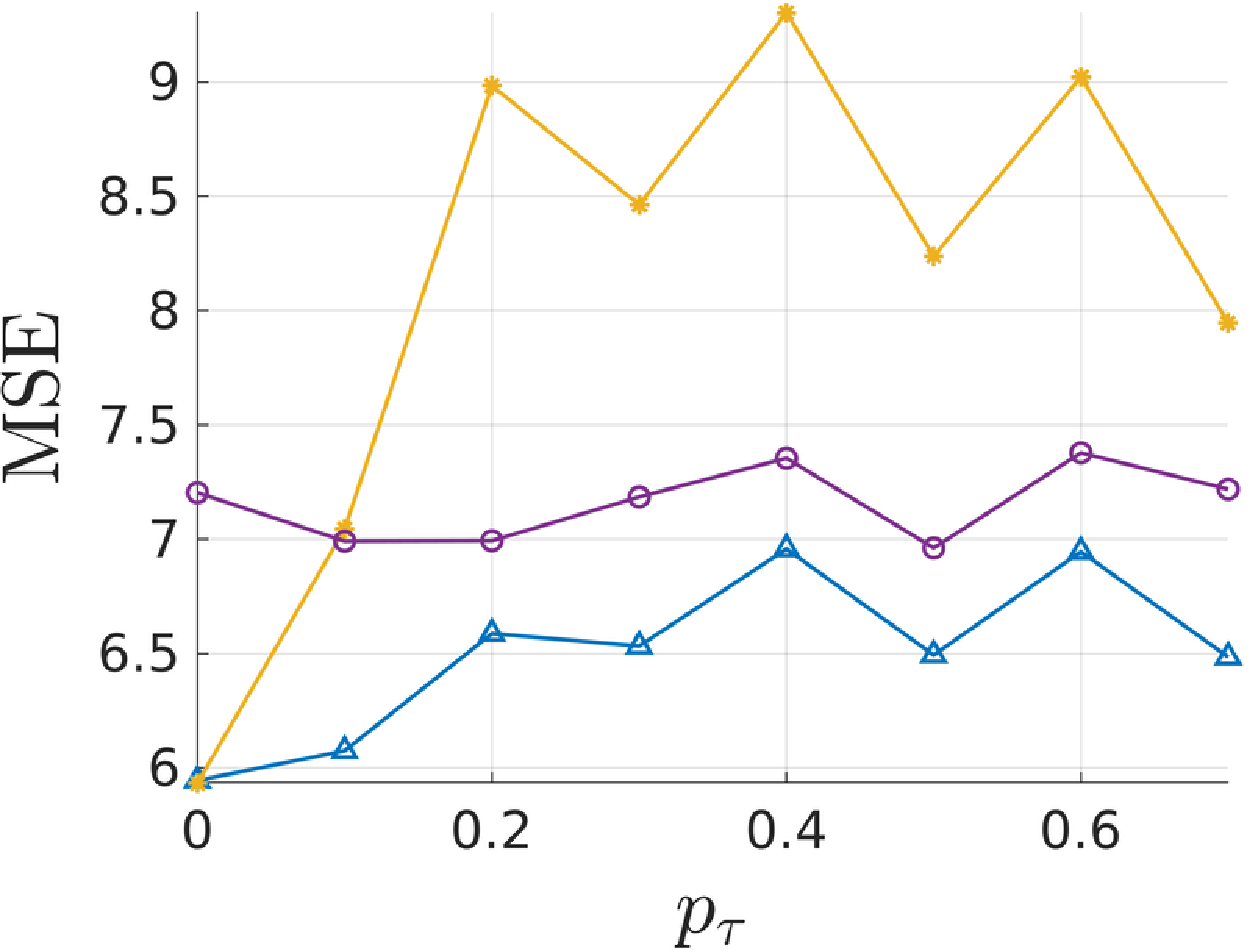}
\centering (b)
\end{minipage}
\hfill
\caption{MSEs of the estimated dense covariance matrices for varying perturbation proportions $p_\tau$ for (a) an MGGD ($\beta=1.5$) and (b) Gaussian distribution ($\beta=2$).}
\label{fig:tau_exp}
\end{figure}

We investigated the different estimator behaviours with varying perturbation levels and fixed $K=20$ and $N=100$. Fig.~\ref{fig:tau_exp} shows the MSEs of the empirical (perturbed when $p_\tau >0$), Tyler's, and proposed dense covariance matrix estimators\footnote{Similar behaviour is observed for the precision matrix estimates, while the mean estimates follow previous findings and are comparable to Tyler's ones for varying perturbation levels.} with different $\tau$ proportions. The experiments are conducted with $\beta=2$ and $\beta=1.5$, corresponding to a Gaussian and heavier-tailed distribution. Note that the empirical statistics are expected to perform ideally in the non-perturbed Gaussian case. As expected, the results show the robustness of the proposed method and Tyler's estimates compared to the empirical statistics when the perturbations increase. We can also see that the proposed method outperforms Tyler's estimates in all cases. Interestingly, the proposed method achieves the ideal performance of the empirical statistics in the non-perturbed Gaussian case ($p_\tau =0$, $\beta=2$). In the non-perturbed heavy-tailed scenario ($p_\tau =0$, $\beta=1.5$), the proposed method outperforms the empirical statistics. In contrast, Tyler's estimates are the least accurate when both distributions are not perturbed.
\begin{table}[h]
\centering
\renewcommand*{\arraystretch}{1.1}
\setlength\tabcolsep{1.5pt}
\caption{MSEs obtained for sparse precision matrix estimation when $K=100$ and $N\ll K$.}
%\resizebox{\linewidth}{!}{%
\begin{tabular}{|l|c||c ||c ||c|} 
\hline
   \multirow{2}{*}{$N$}&  Regularized  & GLASSO & GlassoGaussQn   &Proposed \\
&  Tyler  &  && \\
 \hline
10&  0.80 & 0.77 & 0.76&\textbf{0.64} \\
 \hline
 20& 0.88 & 0.74 & 0.76 &\textbf{0.61}   \\
 \hline
 40& 0.65& 0.48 & 0.48 &\textbf{0.42} \\
 \hline
60& 0.78 & 0.51 & 0.48 &\textbf{0.44} \\
 \hline
80& 0.89 & 0.57 &  0.53 &\textbf{0.44} \\
\hline
% \hline
% N &  20  & 40 & 60 & 80 \\
% \hline
% Regularized Tyler&    &  &  & \\
%  \hline
% GLASSO&   &  &  &  \\
%  \hline
% GlassoGaussQn& &  &  &  \\
%  \hline
% Proposed &  &  &  & \\
% \hline
\end{tabular}%
%}
\label{tab:HD_exp}
\end{table}
%
%---------------------------------
\subsection{High-Dimensional Case: $N\ll K$}
%---------------------------------
% 
For the higher dimensional experiments, we run $N_{\text{MC}}=10^3$ Monte Carlo runs for $K=100$ and $N$-values between $10$ and $80$. We focus on an experimental scenario for which sparse regularization is suitable. A uniformly sparse precision matrix is generated with approximately $90\%$ of the off-diagonal elements equal $0$. The sparse regularization parameters providing the best MSEs are used for all methods, and we use the same values as before for all other parameters of the proposed method. Table~\ref{tab:HD_exp} shows the obtained MSEs for the precision matrices estimated with the regularized Tyler, GLASSO, GlassoGaussQn, and proposed methods. Recall that GLASSO, its robust version GlassoGaussQn, and the proposed method employ a sparse $\ell_1$-regularization of the precision matrix. This could explain the gap in performance between these methods and the regularized Tyler's method. The performance of GLASSO is similar to its robust version GlassoGaussQn, although slightly improved for higher $N$-values. We can see that the proposed method results in the smallest errors for all experiments. 
%$C^{-1}$$ = a_{ij} + \eta \ID$
%
%A visual result is depicted in Fig.~\ref{fig:HD_visu} (figure precision matrix)
%
% %---------------------------------
% \subsection{Effect of the Regularization Parameters}
% %---------------------------------
% %
% \NO{Plot errors as a function of the regularization parameters. Comparison with no regularization. A voir, peut-etre remplacer par plot errors vs tau pourcentage.}
%
%----------------------------------
\section{Conclusion}
%----------------------------------
%
\label{sec:conclusion}
This paper has introduced a novel robust and convex parameter estimation method for the perturbed multivariate Generalized Gaussian distribution. This optimization-based approach judiciously combines a reparametrization of the original likelihood with regularization. Unlike existing methods, theoretical convergence guarantees are obtained for all parameters, and experiments show improvements in performance in various simulation scenarios, including the high-dimensional case. Sparse precision matrix estimation is notably improved compared to regularized robust and non-robust plugin methods. These improvements can be useful for various (high-dimensional) machine learning and graph processing applications. Forthcoming work will investigate such applications in the context of real noisy data. 
% \vfill\pagebreak
% Ajouter les points sur lexist. dun model + non plug in et cvg
%
%-----------------------------------------------
\renewcommand{\thesection}{}
\renewcommand{\thesubsection}{\Alph{subsection}}
\section*{Appendices}
\addcontentsline{toc}{section}{Appendices}
% \appendix
%
%-------------------------------------------------------------------------
\subsection{Proof of Proposition \ref{prop:convex}}\label{Apx:prop_convex}
%-------------------------------------------------------------------------
%
% \NO{Proof consists of showing first that $f$ is a sum of proper lower-semicontinuous convex functions, which makes it a lower-semicontinuous convex function. (use of notion of perspective function and the assumptions made in prop 1). Then, based on assumptions on Q and the form of the fct gmu we show that f is also proper (is domain is not empmty), which completes the proof. (add this in sketch of proof? with the definition of the function f with PHI etc?)}\\
%
%\vskip\baselineskip
%\begin{proof}
Let us define the following functions:
\begin{align}\label{e:defPhi}
&\Phi\colon (\RR^K)^N\times \RR^{N}\colon ((\vv_{n})_{1\le n\le N},\vthe) \mapsto \sum_{n=1}^N \varphi(\vv_{n},\theta_{n}),\\
&\varphi\colon \RR^K\times \RR \colon (\vu,\xi)\mapsto
\begin{cases}
\displaystyle \frac{\|\vu\|^\beta}{2\xi^{\beta-1}} & \mbox{if $\vu \neq \vzero$ and $\xi > 0$}\\
0 & \mbox{if $\vu = \vzero$ and $\xi=0$}\\
\pinf & \mbox{otherwise,}
\end{cases}
\label{e:defvarhi}
\end{align}
and the linear operator
\begin{multline}\label{e:defcalT}
\mathcal{T}\colon \sm \times \RR^{K} \to (\RR^K)^N\colon
(\vQ,\vm) \mapsto  (\vQ \vy_{n}-\vm)_{1\le n \le N}.
\end{multline}
%--old
% \\
% \label{e:defPhi}
% &\Phi\colon (\RR^K)^N\times \RR^K\colon ((\vv_{n})_{1\le n\le N},\vthe) \mapsto \sum_{n=1}^N \varphi(\vv_{n},\theta_{n}),\\
% &\varphi\colon \RR^K\times \RR \colon (\vu,\xi)\mapsto
% \begin{cases}
% \displaystyle \frac{\|\vu\|^\beta}{2\xi^{\beta-1}} & \mbox{if $\vu \neq \vzero$ and $\xi > 0$}\\
% 0 & \mbox{if $\vu = \vzero$ and $\xi=0$}\\
% \pinf & \mbox{otherwise,}\label{e:defvarphi}
% \end{cases}
% \end{align}
% and the linear operator
% \begin{multline}
% \mathcal{T}\colon \sm \times \RR^{K}\times (\RR^K)^N \to (\RR^K)^N\colon\\
% (\vQ,\vm,\vd) \mapsto  (\vQ \vy_{n}-\vm-\vd_{n})_{1\le n \le N}.
Function $\varphi$ is the lower-semicontinuous envelope of the perspective function of $\|\cdot\|^\beta/2$ \cite[Example 9.43]{Livre1}.
Thus, function $\Phi$ defined in \eqref{e:defPhi} belongs to $\Gamma_{0}\big((\RR^K)^N\times \RR^{N}\big)$ and the function
\[
(\vQ,\vm,\vthe) \mapsto  \Phi\big(\mathcal{T}(\vQ,\vm),\vthe\big),
\]
which corresponds to the composition of a linear operator with a proper lower-semicontinuous convex function 
is also proper, lower-semicontinuous, and convex. In addition, 
\begin{equation}
(\forall \vQ \in \sm)\quad \Psi(\vQ) = N \sum_{k=1}^K \psi(\sigma_{k}),
\end{equation}
where $(\sigma_{k})_{1\le k\le K}$ are the eigenvalues of $\vQ$.
%and
%\begin{equation}
%\psi\colon \RR \to \RX\colon \xi \mapsto 
%\begin{cases}
%-\log \xi & \mbox{if $\xi > 0$}\\
%\pinf & \mbox{otherwise.}
%\end{cases}
%\end{equation}
$\Psi$ is a spectral function (\textit{i.e.}, a symmetric function of the eigenvalues of its matrix argument in $\sm$), which is
associated with $\psi \in \Gamma_{0}(\RR)$. Therefore $\Psi \in \Gamma_{0}(\sm)$.

The considered cost function can be re-expressed as
\begin{multline}
\label{e:costreexpressed}
f(\vQ,\vm,\vthe) = \Phi\big(\mathcal{T}(\vQ,\vm),\vthe\big)+ \Psi(\vQ) 
+g_{\mathsf{Q}}(\vQ)\\+g_{\mathsf{m}}(\vm)+\widetilde{g}_{\vartheta}(\vthe)
\end{multline}
and it follows from the previous observations on the two first functions $\Phi\big(\mathcal{T}\cdot,\cdot\big)$ and $\Psi$, and the assumptions made on the three last ones that $f$ is lower-semicontinuous and convex. Then, it is left to show that the domain of $f$ is non-empty, \textit{i.e.,} $f$ is proper.
First, $\emp \neq \dom \widetilde{g}_{\vartheta}$ and,
according to \eqref{e:deftildeg}, 
$\widetilde{g}_{\vartheta}(\vthe)$ is not defined if $\vthe \not \in \RPP^K$,
which means that $\dom \widetilde{g}_{\vartheta} \subset \RPP^N$.
%\NO{according to \eqref{e:deftildeg} (NO: why?)}, $\emp \neq \dom \widetilde{g}_{\vartheta} \subset \RPP^K$ \NO{(NO: $K$ or $N$?)}. 
Since we have assumed that $g_{\mathsf{Q}}$ is also finite at least for one symmetric positive definite matrix, $f$ is proper. Thus, we have shown that $f$ is a proper lower-semicontinuous convex function on $\sm\times \RR^K\times \RR^N$.
%\end{proof}
%
%----------------------------------------------------------------------
\subsection{Proof of Proposition \ref{prop:minim}}\label{Apx:prop_minim}
%----------------------------------------------------------------------
%
We have shown above that $f$ is proper, lower-semicontinuous and convex. Now, we will show that a minimizer exists and provide the conditions for its uniqueness.
\vskip\baselineskip

\begin{proof}
Since$\|\cdot\|^\beta$ with $\beta > 1$  is a strictly convex function, 
for given values of $(\vQ,\vthe)\in \smp\times \RPP^N$,
\begin{equation}\label{e:costm}
\vm \mapsto \Phi\big(\mathcal{T}(\vQ,\vm),\vthe\big)+g_{\mathsf{m}}(\vm)
\end{equation}
is a strictly convex function. Assumption \ref{a:pmini} yields
%Since $g_{\mathsf{m}} \ge 0$,
\begin{multline}
(\forall \vm\in \RR^{K})\quad  \Phi\big(\mathcal{T}(\vQ,\vm),\vthe\big)+g_{\mathsf{m}}(\vm)\\\ge 
 \frac12 \sum_{n=1}^N \frac{\| \vQ \vy_{n}-\vm\|^\beta}{\theta_{n}^{\beta-1}}.
\end{multline}
This allows us to deduce from the coercivity of $\|\cdot\|^\beta$ that the function in \eqref{e:costm} is 
a coercive lower-semicontinuous strictly convex function. Its infimum is thus reached for a unique vector
$\widehat{\vm}(\vQ,\vthe)$. \\

Let function $\Theta$ be defined as
\begin{multline}
(\forall (\vQ,\vthe)\in \sm\times \RR^N)\\
\Theta(\vQ,\vthe)= \inf_{\vm\in \RR^{K}} \Phi\big(\mathcal{T}(\vQ,\vm),\vthe\big)+g_{\mathsf{m}}(\vm)+
\Psi(\vQ)\\+\widetilde{g}_{\vartheta,1}(\vthe).
\end{multline}
It follows from \cite[Proposition 8.35]{Livre1} that the marginal function $(\vQ,\vthe) \mapsto \inf_{\vm\in \RR^{K}} \Phi\big(\mathcal{T}(\vQ,\vm),\vthe\big) + g_{\mathsf{m}}(\vm)$ is convex and from the convexity of the last two terms that $\Theta$ is a convex function.
In addition, by incorporating the infimum $\widehat{\vm}(\vQ,\vthe)$, and by using Definition \eqref{e:defPsi} and Assumption \ref{a:pminii}, for every $(\vQ,\vthe)\in \smp\times \RPP^N$ we have
\begin{multline}
\Theta(\vQ,\vthe) = \Phi\big(\mathcal{T}(\vQ,\widehat{\vm}(\vQ,\vthe)),\vthe\big)
+g_{\mathsf{m}}(\widehat{\vm}(\vQ,\vthe))\\+\Psi(\vQ)+\widetilde{g}_{\vartheta,1}(\vthe)<\pinf.
\end{multline}
This shows that the domain of $\Theta$ is the open set $\smp\times \RPP^N$. It thus follows from
\cite[Corollary 8.39]{Livre1} that $\Theta$ is continuous on this domain.  In addition let $(\check{\vQ},\check{\vthe})\in \smsp\times \RP^N$ be a point on the border of $\dom\Theta$. One of the eigenvalues of $\check{\vQ}$ or one of the components of $\check{\vthe}$ is thus equal to zero. As a consequence of the form of $\Psi$ in \eqref{e:defPsi} and Assumption \ref{a:pminii},
\begin{multline}
\lim_{\substack{(\vQ,\vthe)\to (\check{\vQ},\check{\vthe})\\(\vQ,\vthe) \in \dom \Theta}} \Theta(\vQ,\vthe) 
\ge \\\!\!\lim_{\substack{(\vQ,\vthe)\to (\check{\vQ},\check{\vthe})\\ \vQ \in \smp, \vthe \in \RPP^N}} \!\!\Psi(\vQ)+\widetilde{g}_{\vartheta,1}(\vthe)  = \pinf.
\end{multline}
It thus follows from \cite[Proposition 9.33]{Livre1} that $\Theta$ is a proper lower-semicontinuous convex function.
%Since $\dom f \subset \smp\times \RR^K\times \RPP^N\times (\RR^K)^N$, 
% for every $(\vQ,\vm,\vthe,\vd)\in 
%\smp\times \RR^K\times \RPP^N\times (\RR^K)^N$, 
Besides, 
\begin{align}
\inf_{\vQ\in \sm,\vm \in \RR^K,\vthe\in \RR^N}
&f(\vQ,\vm,\vthe)\nonumber \\
= \inf_{\vQ\in \sm,\vthe\in \RR^N}
& \Theta(\vQ,\vthe)+ g_{\mathsf{Q}}(\vQ)+\widetilde{g}_{\vartheta,0}(\vthe).
%\Phi\big(\mathcal{T}(\vQ,\widehat{\vm}(\vQ,\vthe,\vd),\vd),\vthe\big)+ \Psi(\vQ)+g_{\mathsf{Q}}(\vQ)\nonumber\\
%& +g_{\mathsf{m}}(\widehat{\vm}(\vQ,\vthe,\vd))+\widetilde{g}_{\vartheta}(\vthe)+g_{\mathsf{d}}(\vd)
 \end{align}
 The function
 \begin{equation}\label{e:infcost}
 (\vQ,\vthe) \mapsto \Theta(\vQ,\vthe)+ g_{\mathsf{Q}}(\vQ)+\widetilde{g}_{\vartheta,0}(\vthe)
 \end{equation}
 is lower-semicontinuous and convex as a finite sum of lower-semicontinuous convex functions. It is also proper since, according to the assumptions in Proposition~\ref{prop:convex} and Assumption \ref{a:pminii}, there exists at least one point $(\overline{\vQ},\overline{\vthe})\in \smp\times \RPP^N$ such that
 $g_{\mathsf{Q}}(\overline{\vQ})$, $+\widetilde{g}_{\vartheta,0}(\overline{\vthe})$ is finite.
 In addition, for every $(\vQ,\vthe)\in \sm\times \RPP^N$,
 \begin{multline}
 \Theta(\vQ,\vthe)+ g_{\mathsf{Q}}(\vQ)+\widetilde{g}_{\vartheta,0}(\vthe)
 \ge \\ \Psi(\vQ)+g_{\mathsf{Q}}(\vQ)+\widetilde{g}_{\vartheta}(\vthe).
 \end{multline}
 Using now Assumption \ref{a:pminiii}, we deduce that function \eqref{e:infcost} is coercive. Since it is a coercive proper lower-semicontinuous convex function, it admits a minimizer
$(\widehat{\vQ},\widehat{\vthe})$. Hence, a minimizer of $f$ is 
$\big(\widehat{\vQ},\widehat{\vm}(\widehat{\vQ},\widehat{\vthe}),\widehat{\vthe})$.
In addition, if $g_{\mathsf{Q}}$ and $\widetilde{g}_{\vartheta}$ are strictly convex,
%\NO{(NO: why only $g_{\mathsf{Q}}$ and $\widetilde{g}_{\vartheta}$ ?)}, 
function \eqref{e:infcost} is strictly convex, which ensures that it has a unique minimizer.
\end{proof}
%
%-------------------------------------------
\subsection{Study of Function $\overline{f}$} \label{a:studyfbar}
%-------------------------------------------
%
The first and second-order derivatives of $\overline{f}$ in \eqref{e:ovfopteta} read
\begin{align}
(\forall &\theta_n \in \RPP)\nonumber\\
&\overline{f}'(\theta_{n})= \frac{1}{\theta_{n}}\left(1-
\Big(\frac{\overline{\theta}_n}{\theta_n}\Big)^{\beta-1}
+ \overline{\kappa}(\theta_{n}^{\alpha}-1)\right)\nonumber\\
&\overline{f}''(\theta_{n})
= \frac{1}{\theta_{n}^{2}}\left(
\beta \Big(\frac{\overline{\theta}_n}{\theta_n}\Big)^{\beta-1}
+ (\alpha-1)\overline{\kappa}\theta_n^{\alpha}+\overline{\kappa}-1\right).
\end{align}
The second derivative being positive, $\overline{f}$ is a strictly convex function. Since $\overline{f}(\theta_n) \to +\infty$ as $\theta_n \to 0$ and $\theta_n\to +\infty$, $\overline{f}$ has a unique minimizer. This minimizer is a function of $\overline{\theta}_n$, $\overline{\kappa}$, $\alpha$, and $\beta$ but, for simplicity's sake, we will not make it explicit in our notation. The minimizer $\widehat{\theta}_n$ satisfies the first-order optimality condition
\begin{equation}\label{e:opttheta}
    1-
\Big(\frac{\overline{\theta}_n}{\widehat{\theta}_n}\Big)^{\beta-1}
+ \overline{\kappa}(\widehat{\theta}_{n}^{\alpha}-1) = 0, 
\end{equation}
and we have
\begin{align}
    \overline{f}'(1)&= 1-\overline{\theta}_n^{\beta-1}\\
    \overline{f}'(\overline{\theta}_{n})&= \frac{\overline{\kappa}}{\overline{\theta}_{n}}(\overline{\theta}_{n}^{\alpha}-1).
\end{align}
We deduce that, if $\overline{\theta}_n > 1$, then $\overline{f}'(1)< 0$ and $\overline{f}'(\overline{\theta}_{n}) > 0$, which shows that $\widehat{\theta}_n \in]1,\overline{\theta}_n[$. Similarly, if  $\overline{\theta}_n < 1$, then $\widehat{\theta}_n \in ]\overline{\theta}_n,1[$. In addition, by implicit derivation of \eqref{e:opttheta},
\begin{align}
    & q(\widehat{\theta}_n)
    \frac{\partial \widehat{\theta}_n}{\partial \overline{\theta}_n} = 
    (\beta-1) \overline{\theta}_n^{\beta-2} \widehat{\theta}_n^{1-\beta}\\ 
    & q(\widehat{\theta}_n)
    \frac{\partial \widehat{\theta}_n}{\partial \alpha} = 
    -\overline{\kappa}\widehat{\theta}_n^\alpha \log \widehat{\theta}_n \\
    & q(\widehat{\theta}_n)
    \frac{\partial \widehat{\theta}_n}{\partial \overline{\kappa}} = 1-\widehat{\theta}_n^{\alpha}\\
    &q(\widehat{\theta}_n)\frac{\partial \widehat{\theta}_n}{\partial \beta} = 
    \ln\Big( \frac{\overline{\theta}_n}{\widehat{\theta}_n}\Big)
    \Big(\frac{\overline{\theta}_n}{\widehat{\theta}_n}\Big)^{\beta-1}
\end{align}
where $q(\widehat{\theta}_n) = (\beta-1)\overline{\theta}_n^{\beta-1}\widehat{\theta}_n^{-\beta}
    + \overline{\kappa}\alpha \widehat{\theta}_n^{\alpha-1}$.
Since $\partial \widehat{\theta}_n/\partial \overline{\theta}_n > 0$,
$\widehat{\theta}_n$ is an increasing function of $\overline{\theta}_n$.
If $\overline{\theta}_n > 1$, then $\partial \widehat{\theta}_n/\partial \alpha < 0$, $\partial \widehat{\theta}_n/\partial \overline{\kappa} < 0$,
and $\partial \widehat{\theta}_n/\partial \beta > 0$, which shows that
$\widehat{\theta}_n$ decreases w.r.t. $\alpha$ and 
$\overline{\kappa}$, and increases w.r.t. $\beta$.
Conversely, when $\overline{\theta}_n < 1$, $\widehat{\theta}_n$ is an increasing function w.r.t. $\alpha$ and 
$\overline{\kappa}$, and a decreasing one with respect to $\beta$.

Note that, since $\widehat{\theta}_n$ increases w.r.t. $\overline{\theta}_n$,
$\widehat{\theta}_n \to \widehat{\theta}_{\infty}$ as
$\overline{\theta}_n \to +\infty$, where the limit $\widehat{\theta}_{\infty}$
is either finite or $+\infty$. Eq.~\eqref{e:opttheta} allows us to discard the former case
and to conclude that $\lim_{\overline{\theta}_n \to +\infty}\widehat{\theta}_n = +\infty$.
%
%-------------------------------------------
\subsection{Properties of Operators $\mathbf{L}_1$ and $\mathbf{L}_2$}\label{a:propL1L2}
%-------------------------------------------
%
From the definitions of $\mathbf{L}_1$, $\mathbf{L}_2$, and the norms equipping
the primal and dual spaces $\mathcal{H}$, $\mathcal{G}_1$, and $\mathcal{G}_2$, we deduce that
\begin{align}\label{e:normL1}
\|\mathbf{L}_1\|_{\rm S} &=
\sup_{\mathbf{p}\in \mathcal{H}\setminus\{0\}}
\frac{\|\mathbf{L}_1\mathbf{p}\|_{\mathcal{G}_1}}{\|\mathbf{p}\|_{\mathcal{H}}}=
\max\{\|\mathcal{T}\|_{\rm S},\sqrt{\omega_1}\}\\
\label{e:normL2}
\|\mathbf{L}_2\|_{\rm S} & =\sup_{\mathbf{p}\in \mathcal{H}\setminus\{0\}}
\frac{\|\mathbf{L}_2\mathbf{p}\|_{\mathcal{G}_2}}{\|\mathbf{p}\|_{\mathcal{H}}}= \max\{1,\sqrt{\omega_2}\}.
\end{align}
Let us now evaluate the norm of $\mathcal{T}$.
For every $\vQ \in \mathcal{S}_K$ and
$\vm \in \RR^K$,
\begin{align}\label{e:normT1}
\|\mathcal{T}(\vQ,\vm)\|^2
&= \sum_{n=1}^N \|\vQ \vy_n-\vm \|^2\nonumber\\
&= \sum_{n=1}^N
\Big\| [\vQ\;\,-\vm]\begin{bmatrix}
\vy_n\\
1
\end{bmatrix}
\Big\|^2\nonumber\\
& = \operatorname{tr}([\vQ\;\,-\vm] \mathbf{Y}\mathbf{Y}^\top
[\vQ\;\,-\vm]^\top),
\end{align}
where $\mathbf{Y}$ is given by \eqref{e:defY}. We also have
\begin{align}\label{e:normT2}
&[\vQ\;\,-\vm] \mathbf{Y}\mathbf{Y}^\top
[\vQ\;\,-\vm]^\top\nonumber\\
&\preceq \|\mathbf{Y}\mathbf{Y}^\top\|_{\rm S} [\vQ\;\,-\vm][\vQ\;\,-\vm]^\top\nonumber\\
&=
\|\mathbf{Y}\|_{\rm S}^2 [\vQ\;\,-\vm][\vQ\;\,-\vm]^\top.
\end{align}
Combining \eqref{e:normT1} and \eqref{e:normT2} yields
\begin{align}
\|\mathcal{T}(\vQ,\vm)\|^2
& \le \|\mathbf{Y}\|_{\rm S}^2 \operatorname{tr}([\vQ\;\,-\vm]
[\vQ\;\,-\vm]^\top)\nonumber\\
&= \|\mathbf{Y}\|_{\rm S}^2 
\| [\vQ\;\,-\vm] \|_{\rm F}^2\nonumber\\
& = \|\mathbf{Y}\|_{\rm S}^2
(\|\vQ\|_{\rm F}^2+\|\vm\|^2).
\end{align}
This shows that 
$\|\mathcal{T}\|_{\rm S}\le \|\mathbf{Y}\|_{\rm S}$. Using \eqref{e:normL1} and \eqref{e:normL2}, we deduce that 
\eqref{e:condconvf} is a 
sufficient condition for
\eqref{e:condconv} to be satisfied.\\

The adjoint of $\mathbf{L}_1$ is
\begin{align}
\mathbf{L}_1^*&\colon
\mathcal{G}_1 \to  \mathcal{H}\colon
\big((\mathbf{u}_n)_{1\le n \le N},\vthe)
\mapsto (\vQ,\vm,\omega_1\vthe)
\end{align}
with $(\vQ,\vm)= \mathcal{T}^*\big((\mathbf{u}_n)_{1\le n \le N}\big)$.
The adjoint of $\mathcal{T}$ can be deduced from the identity
\begin{align}
(\forall \vQ \in \mathcal{S}_K)&(\forall
\vm \in \RR^K)(\forall (\mathbf{u}_n)_{1\le n \le N}\in (\RR^K)^N)\\
&\sum_{n=1}^N \mathbf{u}_n^\top [\mathcal{T}(\vQ,\vm)]_n\nonumber\\
&= \sum_{n=1}^N \mathbf{u}_n^\top(\vQ \vy_n-\vm)
\nonumber\\
&= \operatorname{tr}\Big(
\vQ \sum_{n=1}^N\vy_n \mathbf{u}_n^\top\Big)
-\vm^\top \sum_{n=1}^N \mathbf{u}_n,
\end{align}
which shows that
\begin{equation}
\mathcal{T}^*((\mathbf{u}_n)_{1\le n \le N})
= \Big(\frac12\sum_{n=1}^N(\mathbf{u}_n \vy_n^\top+\vy_n \mathbf{u}_n^\top),
-\sum_{n=1}^N \mathbf{u}_n\Big).
\end{equation}
In turn, the adjoint of $\mathbf{L}_2$ is simply expressed as
\begin{equation}
\mathbf{L}_2^*\colon  (\vQ,\vthe)
\mapsto (\vQ,\mathbf{0},\omega_2\vthe).
\end{equation}

%--
\subsection{Proximity Operators}\label{Apx:prox}
%--
The expressions of the proximity operators of the functions involved in \eqref{e:costreexpressed}, up to a positive scaling parameter $\gamma$, are provided below.\footnote{See {http://proximity-operator.net}.}

\begin{itemize}
\item Function $\psi$: For every $\xi \in \RR$,
\begin{equation}
\prox_{\gamma \psi}(\xi) = \frac{\xi+\sqrt{\xi^2+4\gamma}}{2}.
\end{equation}
\item Function $\Psi$: For every $\vQ\in \sm$, let us perform the eigenvalue decomposition of $\vQ$ as
$\vU \Diag(\sigma_{1},\ldots,\sigma_{K}) \vU^\top$ where $(\sigma_{k})_{1\le k \le K}$
are the eigenvalues of $\vQ$ and the columns of $\vU$ form the associated orthonormal basis of eigenvectors. Then,
\begin{multline}
\prox_{\gamma\Psi}(\vQ)\\ = \vU \Diag\big(\prox_{N\gamma \psi}(\sigma_{1}),\ldots,\prox_{N\gamma \psi}(\sigma_{K})\big) \vU^\top.
\end{multline}
%\vskip \baselineskip
%
\item Function $\Phi$: As shown by \eqref{e:defPhi}, $\Phi$ is a separable function of the components of its arguments. 
Thus, 
\begin{multline}
\big(\forall (\vv_{n})_{1\le n\le N} \in (\RR^K)^N\big)(\forall \vthe\in \RR^{N})\quad \\
\prox_{\gamma \Phi} ((\vv_{n})_{1\le n\le N},\vthe)  = \big(\prox_{\gamma \varphi}(\vv_{n},\theta_{n})\big)_{1\le n\le N}.
\end{multline}
                                                                        The expression of the proximity operator of the perspective function $\varphi$ has been derived in \cite[Example~3.7]{Com18}. Let 
\begin{align}
\beta^* = \frac{\beta}{\beta-1}, \qquad \varrho = \left(2\Big(1-\frac{1}{\beta^*}\Big)\right)^{\beta^*-1}
\end{align}
and, for every $\vu \in \RR^K\setminus\{\vzero\}$ and $\xi \in \RR$ such that 
$\beta^* \gamma^{\beta^*-1} \xi + \varrho \|\vu\|^{\beta^*} > 0$, let
\begin{equation}
\vp(\vu,\xi) =
%begin{cases}
%\displaystyle 
\frac{t(\vu,\xi)}{\|\vu\|} \vu,
%& \mbox{if $\vu \neq \vzero$}\\
%0 & \mbox{otherwise,}
%\end{cases}
\end{equation}
where
%, when $\vu \neq \vzero$, 
$t(\vu,\xi)$ is the unique solution in $\RPP$ of the equation
\begin{equation}
s^{2\beta^*-1}+\frac{\beta^*\xi}{\gamma \varrho} s^{\beta^*-1}+\frac{\beta^*}{\varrho^2} s - \frac{\beta^*}{\gamma\varrho^2}\|\vu\|= 0.
\end{equation}
Then, for every $\vu \in \RR^K$ and $\xi\in \RR$,
\begin{align}
&\prox_{\gamma \varphi}(\vu,\xi) =\\
&\begin{cases}
\displaystyle \Big(\vu-\gamma \vp(\vu,\xi) ,\xi+\frac{\gamma\varrho}{\beta^{*}} t(\vu,\xi)^{\beta^*}\Big) & \mbox{if $\vu \neq \vzero$ and } \\
&\mbox{\hspace{-2.5em}$\beta^* \gamma^{\beta^*-1} \xi + \varrho \|\vu\|^{\beta^*} > 0$}
% or $\xi >0$}
\\
(\vzero,\xi) & \mbox{if $\vu = \vzero$ and $\xi > 0$}\\
(\vzero,0) & \mbox{otherwise,}
\end{cases}\nonumber\\
&=  \big(\prox_{\gamma \varphi}^{(1)}(\vu,\xi),\prox_{\gamma \varphi}^{(2)}(\vu,\xi)\big).
\end{align}
As shown by the developments in Section~\ref{se:PFalgo} (see \eqref{eq:w1} -- \eqref{eq:w2}),
one may be interested in deriving the expression of the proximity operator in a metric that weights the two variables in an unbalanced manner. 
%\NO{Why and when useful exactly?} 
This means that, for every $(\gamma_{1},\gamma_{2})\in [0,+\infty[^2$, $\vu \in \RR^K\setminus\{\vzero\}$ and $\xi \in \RR$, one would seek
\begin{multline}
\prox_{\varphi}^{\gamma_{1},\gamma_{2}}(\vu,\xi)
= \argmind{\substack{\vu\in \RR^K,\\\xi'\in \RP}}\!\! \varphi(\vu',\xi') + \frac{1}{2\gamma_{1}}\|\vu'-\vu\|^{2}\\+\frac{1}{2\gamma_{2}} (\xi'-\xi)^2.
\end{multline}
Performing the variable change $\tilde{\vu} = \vu'/\sqrt{\gamma_{1}}$ and $\tilde{\xi} = \xi'/\sqrt{\gamma_{2}}$
and using the definition of $\varphi$ in \eqref{e:defvarhi} yields
\begin{multline}
\prox_{\varphi}^{\gamma_{1},\gamma_{2}}(\vu,\xi)
= \Bigg(\sqrt{\gamma_{1}} \prox_{\gamma \varphi}^{(1)}\Big(\frac{\vu}{\sqrt{\gamma_{1}}},\frac{\xi}{\sqrt{\gamma_{2}}}\Big),\\
\sqrt{\gamma_{2}} \prox_{\gamma\varphi}^{(2)}\Big(\frac{\vu}{\sqrt{\gamma_{1}}},\frac{\xi}{\sqrt{\gamma_{2}}}\Big) \Bigg),
\end{multline}
with $\gamma = \sqrt{\gamma_{1}^{\beta}/\gamma_{2}^{\beta-1}}$.
\end{itemize}

% \AtEndEnvironment{thebibliography}{
% % \begin{thebibliography}{99} 

% }

\bibliographystyle{IEEEbib}
\bibliography{biblio_icassp_2022.bib}

\end{document}